\documentclass{aa}
\usepackage{txfonts} 
\usepackage{times}
\usepackage{graphicx}
\usepackage{natbib} 
\usepackage{color}
\usepackage{upgreek}
\usepackage{xspace}
\usepackage{tipa}
\usepackage{graphicx}

\newcommand{\VVV}{\textcolor{black}{$\surd$}}

\title{Location and sizes of forsterite grains in protoplanetary disks}
\subtitle{Interpretation from the Herschel DIGIT programme}
\author{K. M. Maaskant\inst{1} \and B.L. de Vries\inst{2,3} \and M. Min\inst{4}  \and L.B.F.M. Waters\inst{5,}\inst{4} \and C. Dominik\inst{4,}\inst{6} \and F. Molster \inst{7,8} \and A.G.G.M. Tielens\inst{1} }
\institute{     Leiden Observatory, Leiden University, P.O. Box 9513, 2300 RA Leiden, The Netherlands \and 
                AlbaNova University Centre, Stockholm University, Department of Astronomy, SE-106 91 Stockholm, Sweden               \and
                Stockholm University Astrobiology Centre, SE-106 91 Stockholm, Sweden \and
                Anton Pannekoek Astronomical Institute, University of Amsterdam, P.O. Box 94249, 1090 GE Amsterdam, The Netherlands \and
                SRON Netherlands Institute for Space Research, Sorbonnelaan 2, 3584 CA Utrecht, The Netherlands \and
                Department of Astrophysics/IMAPP, Radboud University Nijmegen, PO Box 9010 6500 GL Nijmegen, The Netherlands   \and
                NOVA, P.O. Box 9513, 2300 RA Leiden, The Netherlands \and
                Dutch Space BV, P.O. Box 32070, 2303 DB Leiden, The Netherlands }

\abstract{The spectra of protoplanetary disks contain mid- and far- infrared emission features produced by forsterite dust grains.  The spectral features contain information about the forsterite temperature, chemical composition and grain size.  }
{We aim to characterize how the 23  and 69 $\upmu$m features can be used to constrain the physical locations of forsterite in disks. We check for consistency between two independent forsterite temperature measurements: the $I_{23}/I_{69}$ feature strength ratio and the shape of the 69 $\upmu$m band.}
{We performed radiative transfer modeling to study the effect of disk properties to the forsterite spectral features. Temperature-dependent forsterite opacities were considered in self-consistent models to compute forsterite emission from protoplanetary disks.  }
{Modelling grids are presented to study the effects of grain size, disk gaps, radial mixing and optical depth to the forsterite features. Independent temperature estimates derived from the $I_{23}/I_{69}$ feature strength ratio and the 69 $\upmu$m band shape are most inconsistent for HD\,141569 and Oph IRS 48. A case study of the disk of HD\,141569 shows two solutions to fit the forsterite spectrum. A model with T $\sim40$ K, iron-rich ($\sim0-1$ \% Fe) and 1 $\upmu$m forsterite grains, and a model with warmer (T $\sim$ 100 K), iron-free, and larger (10~$\upmu$m) grains.}
{We find that for disks with low upper limits of the 69 $\upmu$m feature (most notably in flat, self-shadowed disks), the forsterite must be hot, and thus close to the star. We find no correlation between disk gaps and the presence or absence of forsterite features. We argue that the 69 $\upmu$m feature of the evolved transitional disks HD\,141569 and Oph IRS 48 is most likely a tracer of larger (i.e. $\gtrsim10$ $\upmu$m) forsterite grains. }


\date{\today} 

\begin{document}

\maketitle


\section{Introduction}
Protoplanetary disks are thought to be the precursors of planetary systems. Disk evolution is therefore of great importance for understanding how planets are being formed (e.g. \citealt{2014Testi}). Through analysing the dust emission features at infrared (IR) wavelengths, both the chemical composition of dust and the size of grains can be constrained (e.g. \citealt{2010Henning}). Characterising the dust provides evidence of physical and chemical processes in the disk. 

Silicate dust undergoes significant processing in protoplanetary disks. Whereas dust in the interstellar medium (ISM) is largely amorphous \citep{2004Kemper}, that is, it is characterised by a disordered network of silicates that are enriched with O, Fe, and Mg, and to a lesser degree Ca and Al. In protoplanetary disks, silicate dust is also observed in the crystalline phase \citep{1998Malfait, 2001Bouwman, 2001Meeus}. Forsterite (Mg$_2$SiO$_4$) is the most abundant observable crystalline constituent in  disks around T Tauri and Herbig Ae/Be stars \citep{2009Olofsson, 2010Juhasz}. Forsterite formation requires temperatures of $\gtrsim1000$ K \citep{2000Hallenbeck, 2000Fabian}. At these temperatures, crystalline silicates can be formed through gas-phase condensation and annealing of preexisting amorphous silicates (see \citealt{2005Wooden} for an overview). The required temperatures indicate that forsterite forms in the inner disk. High crystallinity fractions in the inner disk have been confirmed by interferometric observations \citep{2004Boekel}. 

Even though dust crystals are formed at high temperature and therefore close to the star, they have also been observed much farther away from the star than expected \citep{1998Malfait,2009Olofsson, 2010Juhasz, 2012deVries, 2013Sturm}. In our own solar system, comets and interplanetary dust particles also include crystalline silicates \citep{1987MacKinnonRietmeijer, 1992Bradley}. Silicate grains in long-period comets such as Hale-Bopp can be crystalline \citep{1997Crovisier, 1999Wooden, 2002Harker}. Laboratory measurements of olivine crystals from unequilibrated bodies (e.g. such as comet 81P/Wild 2, \citealt{2006Zolensky}; and cometary interplanetary dust particles \citealt{2008Zolensky}) are generally iron poor. Iron fractions (by number) in olivine crystals found in the Itokawa asteroid and in ordinary chondrites can be as high as $\sim$ 30\% \citep{2011Nakamura}. This may reflect the lack of large-scale mixing \citep{1994Shu, 2002Bockelee-Morvan, 2004Gail, 2014Jacquet} and indicates the importance of crystal formation (i.e. local heating events) associated with the inner and outer disk region \citep{1967Urey,2001Huss, 2002HarkerDesch, 2005Desch, 2009Abraham,2010Morlok, 2010Tanaka}.

The transition from amorphous to crystalline silicate dust leads to strong changes in the optical response function of the material \citep{1998Hallenbeck}. Crystalline silicates have several sharp solid-state features. A detailed analysis of these features provides a wealth of information on the properties of the crystal grains \citep{2003Koike, 2006Suto}. Because these properties can potentially be linked to the specific circumstances required to crystallise the dust grains, crystalline silicates allow us to probe the
physical processes that change the chemical compositions in the disk.

The width and peak position of the 69 $\upmu$m feature of forsterite is sensitive to the temperature \citep{2002Bowey, 2006Suto, 2006Koike}. In addition, the peak position of the feature becomes red-shifted for higher fractions of iron \citep{1993Koike, 2003Koike}. Herschel observations using the Photodetector Array Camera and Spectrometer (PACS) have allowed us to study the 69 $\upmu$m feature with high spectral resolution \citep{2010Sturm, 2011Mulders,2012deVries, 2013Sturm}. A detailed analysis of the 69 $\upmu$m feature shapes in protoplanetary disks is presented in \citet{2013Sturm}. These authors found the following results. Most of the forsterite grains that give rise to the 69 $\upmu$m bands are found to be $\sim100-200$ K and iron-poor (less than $\sim$2\% iron), which supports the hypothesis that the forsterite grains form through an equilibrium condensation process at high temperatures. The large width of the emission band in some sources has been interpreted to indicate the presence of forsterite reservoirs at different temperatures. It was also concluded that any model that can explain the PACS and the Spitzer IRS observations must take the effects of a wavelength-dependent optical depth into account. 

Radiative transfer modelling has only been performed for HD\,100546 \citep{2011Mulders} to study the consistency between the relative mid-infrared band strengths and the information contained in the shape of the 69 $\upmu$m band. It was found that the best solution is given by a model with a high abundance of forsterite located in the inner edge of the outer disk. Because protoplanetary disks are highly optically thick media, radiative transfer modelling is essential for investigating optical-depth effects and for simultaneously studying the effects of grain growth, protoplanetary gaps and, radial mixing. 

In this paper, we perform an in-depth study of the 69 $\upmu$m feature of Herbig Ae/Be stars using radiative transfer models. We find that forsterite shows a variety of spectral characteristics for different groups of protoplanetary disks and that the detection rate of the 69 $\upmu$m feature increases for lower disk masses (Sect. \ref{sec:observations}). Because the different groups can be linked to physical processes in the disk, this shows that forsterite is a valuable tracer of the disk evolution history. In Sect. \ref{sec:modeling_approach} we introduce the radiative transfer code and the dust model. We study the behaviour of forsterite in different evolutionary scenarios in Section \ref{sec:relativestrengths}. In Sect. \ref{sec:hd141569} we present a case study of the peculiar 69 $\upmu$m band that may be indicative for large grains originating in the optically thin disk of HD\,141569. The discussion and conclusions are presented in Sects. \ref{sec:discussion} and \ref{sec:conclusions}. 

          \begin{figure}[t] 
\includegraphics[width=\columnwidth]{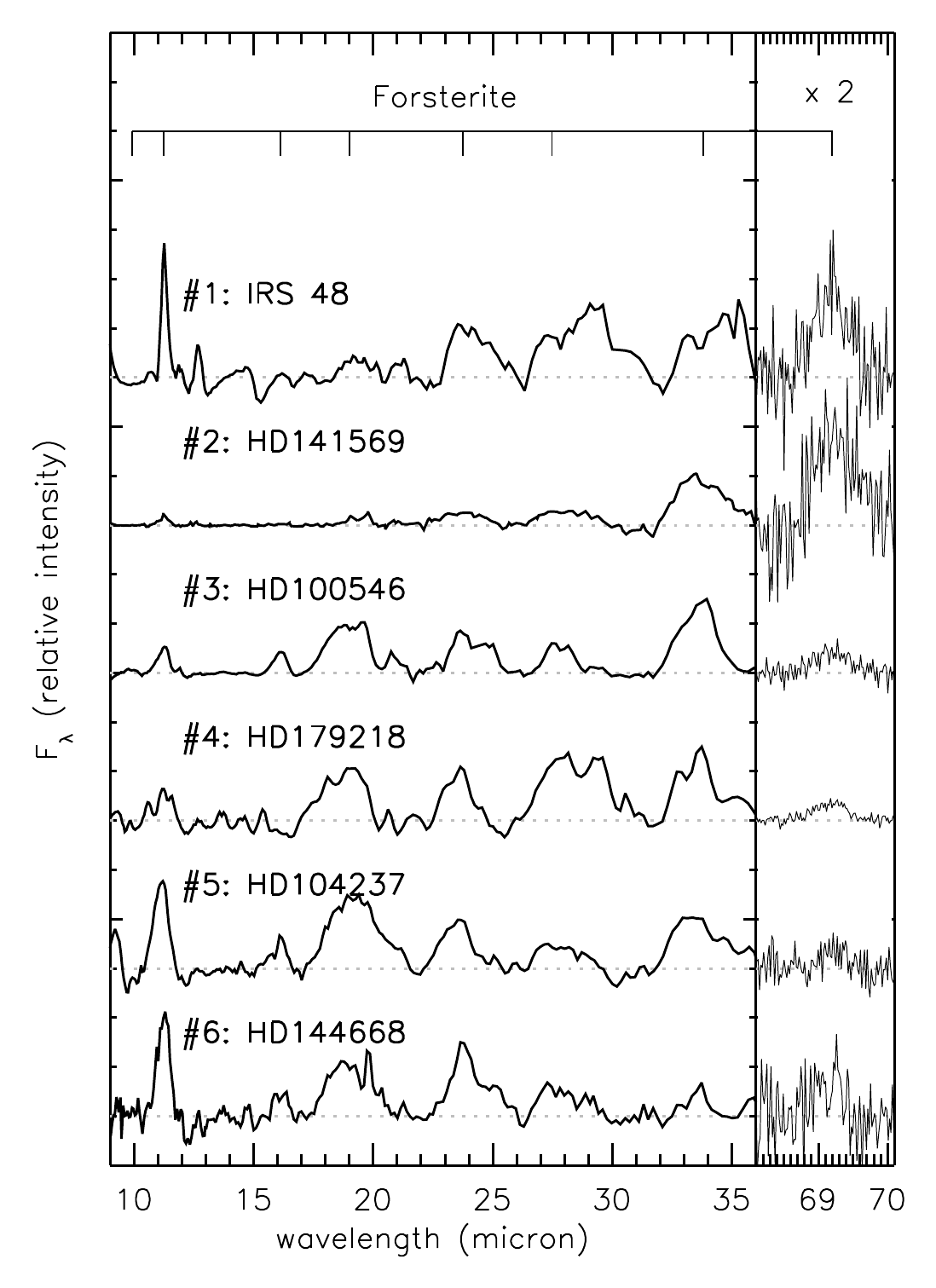}
\caption{\label{fig:detections}  Spectra of all sources with 69 $\upmu$m detections. The continuum-subtracted spectra are scaled relative to the strongest band. The objects are sorted by the $F_{30}/F_{13.5}$ ratio. } 
\end{figure}

\begin{figure}[t] 
\includegraphics[width=\columnwidth]{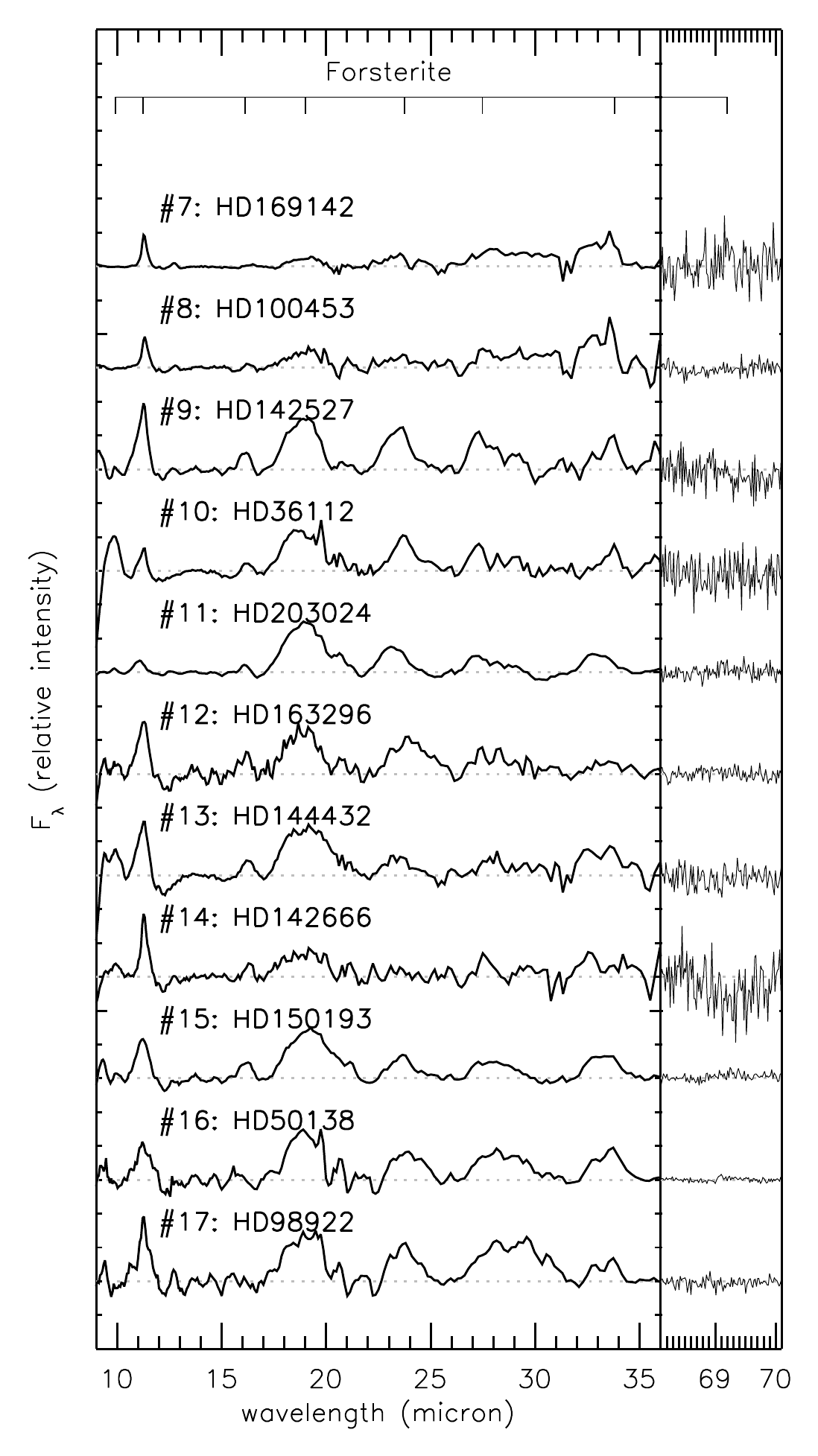}
\caption{\label{fig:non_detections}  Spectra of all sources with three or more forsterite features detected in the Spitzer/IRS spectrum. These objects have no detections at 69 $\upmu$m. The continuum-subtracted spectra are scaled relative to the strongest band and the objects are sorted by the $F_{30}/F_{13.5}$ ratio.} 
\end{figure}


\section{Observations}
\label{sec:observations}
In this section, we analyse the short- and long-wavelength forsterite features originating from protoplanetary disks. We compare the feature strength ratio $I_{23}/I_{69}$ with the forsterite 69 $\upmu$m shape and check for their consistency. In addition, we present trends between the $I_{23}/I_{69}$ ratios and the millimetre (mm) luminosities of the disks. Note that throughout this paper, the integrated intensities are given by $I_{\lambda}$ [erg~cm$^{-2}$~s$^{-1}$], the monochromatic flux by $F_{\lambda}$ [erg~cm$^{-2}$~s$^{-1}$~$\upmu$m$^{-1}$], and the luminosity by $L_{\lambda}$~[erg~s$^{-1}$].

\subsection{Data collection}
We selected far-infrared (FIR) spectroscopic observations of Herbig stars (i.e. stars of spectral type F and higher) from the sample of \citet{2013Sturm}. This allows us to identify trends in a sample of objects that are similar in (stellar) properties. The observations were taken with the PACS instrument and are analysed as part of the Dust, Ice and Gas in Time (DIGIT) Herschel key programme. 
The data were processed using the Herschel Interactive Processing Environment (HIPE), calibration version 42, and standard pipeline scripts. Table \ref{tab:featureshapes} shows the properties of the objects with forsterite 69 $\upmu$m detections. Table \ref{tab:sample} shows all Herbig Ae/Be objects that were observed. For comparison with the spectral energy distributions (SEDs) we collected fluxes at 13.5 and 30 $\upmu$m from \citet{2010Acke} and mm photometry for all sources from the literature. We identified forsterite features in the wavelength range between $\sim15-35$ $\upmu$m using Spitzer/IRS spectra, which were first published by \citet{2010Juhasz} and re-reduced by the most recent calibration version (S18.18.0) for the study by \citet{2013Sturm}.

Identification and measurements of the forsterite features were carried out after fitting and subtracting the continuum. The best fit to the continuum was obtained using a spline function, which is a polynomial between continuum wavelengths chosen just outside the features. The best results were obtained using the following wavelengths for the continuum: for the Spitzer/IRS wavelength domain: $\lambda_{\rm continuum} =$ \{5.52, 5.83, 6.66, 7.06, 9.54, 10.32,12, 13.08, 13.85, 15.00, 17.00, 21.75, 26.00, 31.60, 35.10, and 36.01\} $\upmu$m. For the Herschel/PACS wavelength domain:   $\lambda_{\rm continuum} =$ \{65.50, 67.50, 68.55, 69.70, 70.00, 70.50, 71.00, and 71.50\} $\upmu$m. At each point we took the average continuum flux for a narrow wavelength range ($\Delta \lambda\sim1 \upmu$m) around the feature. An extended discussion of this approach is given in Appendix B of \citet{2011Mulders}. We visually checked that the continuum was properly fitted so that the strengths of the forsterite features could be accurately determined. We checked the spectra for detections of forsterite features. Almost all of the forsterite features we identify are also reported in \citet{2010Juhasz} and \citet{2013Sturm}. Table \ref{tab:sample} gives an overview of all forsterite detections (for the short-wavelength features, this implies that at least three bands are identified). Finally, the 23 $\upmu$m feature was identified using a $\chi^2$ Gaussian fitting routine and thereafter integrated to determine the band strengths. From all the short-wavelength features we chose the 23 $\upmu$m band for analysis because 1) the band is relatively unaffected by the amorphous silicate feature, 2) it is a strong/prominent band, 3) there is minimal interference of other crystalline silicate bands (most notably enstatite), and 4) the Spitzer/IRS observations are reliable at that wavelength range. The error on the flux of the 23 $\upmu$m band is dominated by the choice of the continuum points next to the features.  This error calculation varies the position of the continuum points by a Monte Carlo based method and determines the 1 $\sigma$ deviation to the mean flux. We derive a typical error of $\sim$20\% to the feature strength.

\citet{2013Sturm} report the detection of the 69 $\upmu$m feature in the PACS spectrum of AB Aur. However, several indications make the detection uncertain. The identification of the continuum between $\sim 65-75$ $\upmu$m is arbitrary since the continuum of AB Aur shows several bumps in this wavelength range. Because
of instrumental problems with PACS, it is difficult to identify for the indicated wavelength range whether bumps are instrumental artefacts, real features, or continuum. After analysing the 69 $\upmu$m the shape, the reported forsterite feature has a peculiar square-like shape, which is not seen in the other detections of the Herbig stars. In addition, the peak position is red-shifted by 0.8 $\upmu$m, while the other detections are all within a range of 0.2 $\upmu$m. While this could be an effect of an iron fraction of $\sim$4 \%, it is somewhat peculiar that AB Aur has this iron fraction, and the other detections not. Finally, unlike all other Herbig stars where the 69 $\upmu$m feature is detected, AB Aur has no sign of forsterite detections in the Spitzer/IRS spectrum. Therefore, we chose not to classify the 69 $\upmu$m feature of AB Aur as a detection of forsterite in our analysis. We emphasize that AB Aur is an outlier and better observations are needed to confirm the detection with sufficient confidence.

\subsection{New Meeus group classificÞcation based on the $F_{30}/F_{13.5}$ continuum flux ratio }
\begin{figure}[t] 
\centering
\includegraphics[width=0.5\textwidth]{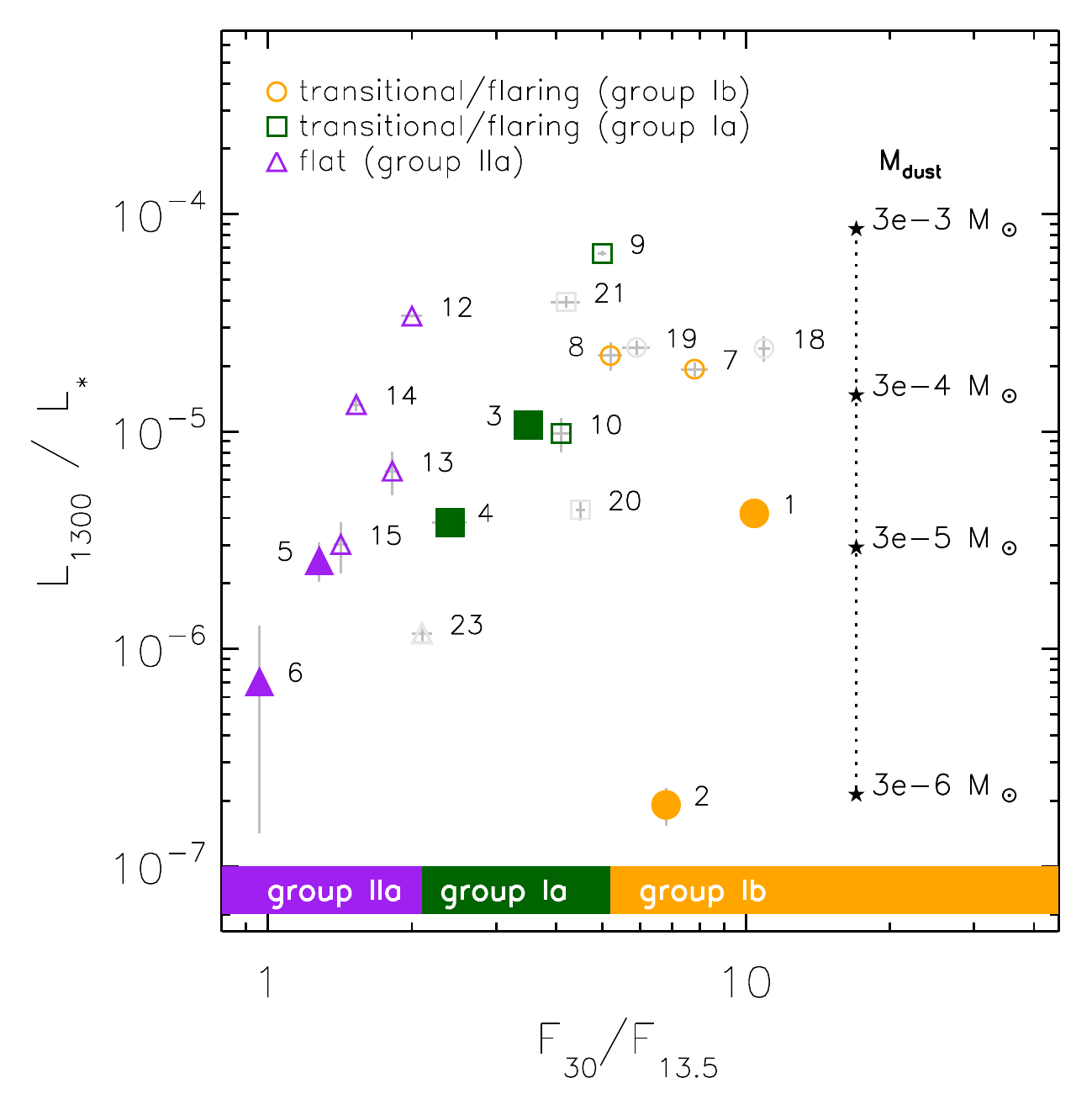}

\caption{\label{fig:L_vs_ratio1} Disk luminosity at 1.3 mm ($L_{1300}/L_{*}$) compared with the MIR spectral index $F_{13.5}/F_{30}$ ratio.  Filled symbols indicate the objects for which the 69 $\upmu$m feature has been detected. The 69 $\upmu$m feature is often detected for disks with low  $L_{1300}/L_{*}$. The MIR spectral index can be used as a tracer of large disk gaps. The symbols show flaring/transitional disks without silicate features (group Ib, orange circles,  $F_{30}/F_{13.5} \geqslant 5.1$), with silicate features (group Ia, dark green squares, $2.1 < F_{30}/F_{13.5} < 5.1 $), and self-shadowed disks with silicate features (group IIa, purple triangles, $F_{30}/F_{13.5} \leqslant 2.1$). The numbers refer to the objects and are given in Table \ref{tab:sample}. The grey symbols show the objects without forsterite detections at Spitzer/IRS and Herschel/PACS wavelengths.} 
\end{figure}

\citet{2001Meeus} showed that the SEDs of Herbig stars can be divided into two typical groups: sources with a strong mid- to far-infrared excess (group I) and sources that can be fitted by just one power law (group II). A strong correlation between the $F_{30}/F_{13.5}$ continuum flux ratio and the group classification based on \citet{2001Meeus} has been shown in \citet{2013Maaskant,2014Maaskant}. Group I and II sources have high and low $F_{30}/F_{13.5}$ ratios,
respectively, and are easily distinguished by this parameter. Larger sizes of the gap in the disk temperature regime between $\sim200-500$K causes the continuum flux at 13.5 $\upmu$m to decrease. However, because the inner edge of the outer disk is a vertical wall with a high surface brightness, the flux at 30 $\upmu$m increases. Therefore high values of $F_{30}/F_{13.5}$ are indicative of large gap sizes \citep{2013Maaskant}. As proposed by Khalafinejad (in prep.), the transitions between Meeus groups Ia, Ib and group IIa can be more easily classified using the $F_{30}/F_{13.5}$ ratio, with geometrically flat, full disks showing amorphous silicate features (group IIa): $F_{30}/F_{13.5} \leqslant 2.1$. For transitional/flaring objects with amorphous silicate features (group Ia): $2.1 < F_{30}/F_{13.5} < 5.1 $. The transitional/flaring objects without silicate features (group Ib) have $F_{30}/F_{13.5} \geqslant 5.1$. The devision of these groups can be seen in Fig. \ref{fig:L_vs_ratio1}. The absence of the silicate feature is correlated with the highest values of $F_{30}/F_{13.5}$ because the disk gap is significantly depleted in small silicate dust grains. Because of the depletion of small dust grains, the optical depth decreases as well, which enhances the flux from smaller PAH molecules that are assumed to be coupled with the gas \citep{2014Maaskant}. For our study we adopted this new classification scheme of the Meeus groups based on the $F_{30}/F_{13.5}$ ratio.

\begin{figure}[t] 
\includegraphics[width=0.5\textwidth]{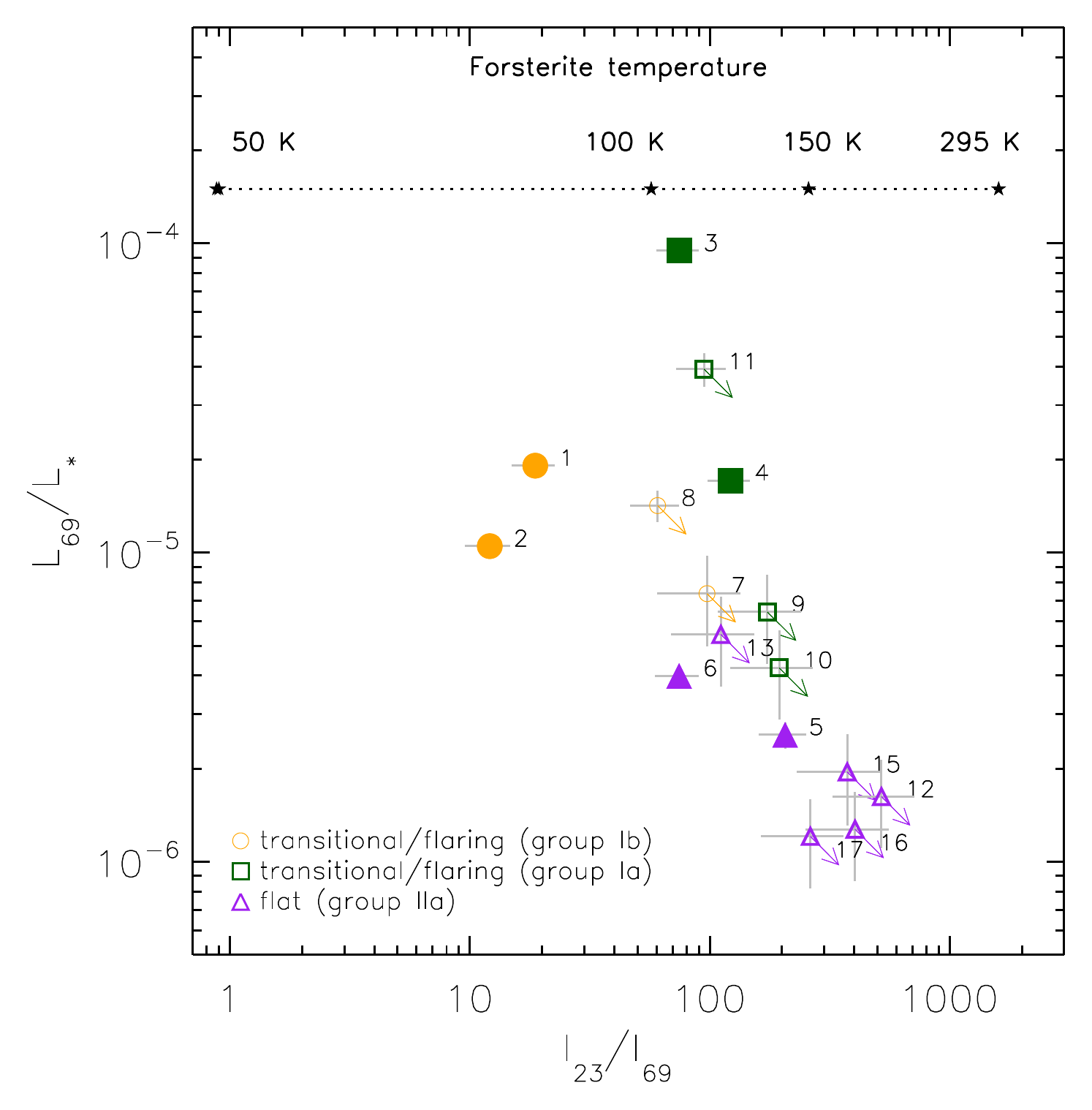}
\caption{\label{fig:69_vs_forstT} Luminosity of the 69 $\upmu$m feature ($L_{69}/L_{*}$) compared with the $I_{23}/I_{69}$ ratio. Filled symbols indicate the objects for which the 69 $\upmu$m feature has been detected. The numbers refer to the objects and are given in Table \ref{tab:sample}. The arrows indicate the upper limits of the 69 $\upmu$m feature. The stars indicate band-strength ratios of temperature-dependent forsterite opacities multiplied by black bodies of corresponding temperatures. Thus the emission profile of optically thin dust emission is modelled and serves as a simple first comparison to the observed feature ratios. } 
\end{figure}

\subsection{Properties of the forsterite features}

Out of 23 sources observed with Herschel/PACS, 6 objects have a detection of the 69 $\upmu$m feature. Their continuum-subtracted Spitzer/IRS and Herschel/PACS spectra are shown in Fig. \ref{fig:detections}. All objects without 69 $\upmu$m feature detections are shown in Fig. \ref{fig:non_detections}. In these figures, the continuum-subtracted spectra are scaled relative to the strongest forsterite band, and the objects are sorted by the continuum-feature-strength ratio of the 30 $\upmu$m and 13.5 $\upmu$m features (i.e. the $F_{30}/F_{13.5}$ ratio). We were unable to identify any parameters that might explain the lack of forsterite features for several sources  (sources \#18 -- \#23, see Table \ref{tab:sample}).

\subsection{Observational trends}

Since we only have very few 69 $\upmu$m detections, extracting trends is difficult, and we need to be careful not to draw strong conclusions. Nevertheless, we searched for trends between the forsterite features and the stellar and photometric properties. The following tentative trends, related to the photometric properties (Fig. \ref{fig:L_vs_ratio1}) and to the forsterite features (Fig.  \ref{fig:69_vs_forstT}), were identified in the data.

Figure \ref{fig:L_vs_ratio1} shows all the Herbig stars that
were observed with Spitzer/IRS and Herschel/PACS. This plot relates the 69  m detections to the photometric properties of the Herbig stars (where 69 $\upmu$m detections are indicated by the filled symbols). One of the main indicators of the protoplanetary disk mass is the 1.3 mm optically thin dust continuum emission. For comparison we have over-plotted mm luminosities of a radiative transfer modelling grid were the mass is varied by four orders of magnitude (Sect. \ref{sec:opt_depth}). This plot shows that the detection rate of the 69 $\upmu$m forsterite features tends to be higher for lower 1.3 mm luminosities ($L_{1300}/L_{*}$). Thus for lower mass disks, the 69 $\upmu$m feature seems to be more frequently detected. We discuss in Sect. \ref{sec:relativestrengths} which disk evolutionary scenarios can explain this behaviour. Figure \ref{fig:L_vs_ratio1} also shows that there is no correlation between the 69 $\upmu$m detections and the presence of large gaps, as indicated by the MIR spectral index $F_{30}/F_{13.5}$ ratio. 

Figure \ref{fig:69_vs_forstT} shows all the sources with forsterite detections in the Spitzer/IRS and/or Herschel/PACS wavelength range. It relates the 69 $\upmu$m luminosity to the Meeus groups as well as to the forsterite temperature. Note that the absolute 69 $\upmu$m feature strength \textit{alone} is a difficult diagnostic tool to search for processes in disks because the abundance differences can have a strong effect on the feature and there is a large dispersion in the crystallinity fractions of Herbig stars.  The  $I_{23}/I_{69}$ band strength \textit{ratio} can be interpreted as a probe of the dominant forsterite temperature in the disk because it measures the shape of the underlying blackbody. Also overlaid are feature strength ratios for optically thin representative dust mixtures of different temperatures (20\% forsterite, 20\% amorphous carbon, and 60\% amorphous silicates, see Sect. \ref{sec:dustmodel} for the description of the dust model). In this mixture, temperature-dependent opacities of forsterite are adopted \citep{2006Suto}. Hence, a high $I_{23}/I_{69}$ ratio reflects a high forsterite temperature.  We can learn from this plot that the luminosity of the 69 $\upmu$m detections is stronger for transitional/flaring disks (group I) compared with the self-shadowed/flat (group II) objects. In addition, the forsterite temperature indicator $I_{23}/I_{69}$ seems to be $\sim1-2$ orders of magnitude lower for group Ib objects (Oph IRS 48 and HD\,141569)  than for  group Ia and IIa objects. The weakest detections of the 69 $\upmu$m feature are found for the flat (group II) objects and have a similar $I_{23}/I_{69}$ ratio as group Ia objects. In the majority of objects, the 69 $\upmu$m feature is not observed. The non-detection could be a consequence of insufficient sensitivity (such as for object \#11: HD\,203024). However, for the flat (group II) objects in the bottom-right corner of Fig. \ref{fig:69_vs_forstT} (\#12, \#15, \#16 and \#17), the 23 $\upmu$m feature is strong compared with the upper limit of the 69 $\upmu$m feature. For these objects we can conclude that the forsterite temperature must be high.

The possible effects of these trends on the relative band strength $I_{23}/I_{69}$ and the 69 $\upmu$m width and peak position are carefully examined in the remainder of this paper.

\subsection{Consistency between feature strength ratio $I_{23}/I_{69}$  and the 69 $\mu$m band shape}
\label{sec:consistency}
Another independent way of estimating the forsterite temperature relies on the analysis of the 69 $\upmu$m feature shape. The feature width and peak position contains information about the temperature, iron fraction, and grain size. We now examine the consistency between the two methods to derive a forsterite temperature estimate.

\begin{figure}[t] 
\includegraphics[width=\columnwidth]{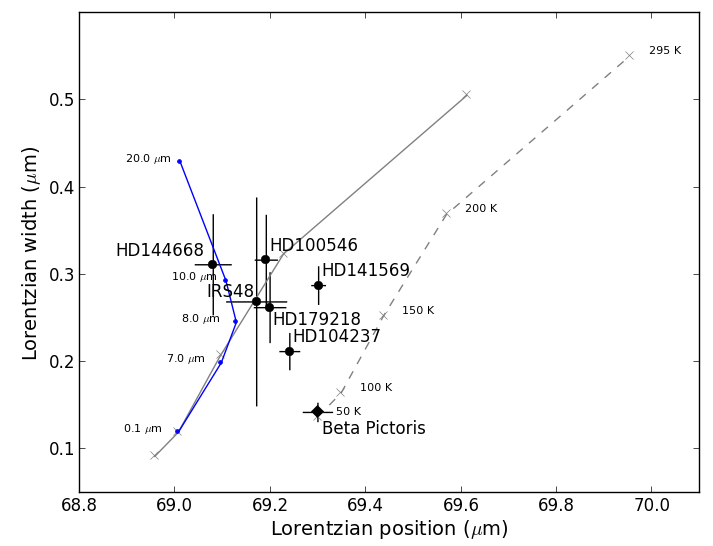}
\caption{\label{fig:deVriesdiagram} Lorentzian width versus peak position diagram of 69 micron features (Gaussian values in Table \ref{tab:featureshapes}). The left solid blue line shows the peak positions and widths for a T = 100 K grain with increasing size between 0.1 $\upmu$m and 20 $\upmu$m, computed with a distribution of hollow spheres (DHS). The left solid grey line represents the peak positions and widths for iron free forsterite and increasing temperatures of 50 K, 100 K, 150 K, 200 K, and 295 from laboratory measurements of \citet{2006Suto}. The right dashed grey line shows the interpolated shift in peak positions and widths for the 69 $\upmu$m feature for 1\% iron fraction \citep{2003Koike}.}

\end{figure}

The feature strengths, widths, and peak positions of the 69 $\upmu$m feature are examined in \citet{2013Sturm}. The feature properties are shown in Table \ref{tab:featureshapes}. Lorentzian fits are also derived for the features to be able to compare them to the detection of the 69 $\upmu$m feature in Beta Pictoris presented in \citet{2012deVries}. The forsterite in the debris disk of Beta Pictoris was found to be $\sim$80 K at a distance of $\sim$40 AU, and contains 1\% iron. Because gas-rich disk may eventually evolve into debris disks, a comparison between the properties of the 69 $\upmu$m feature in Herbig stars to that of Beta Pictoris may be relevant. The feature widths and positions for the sources in our sample are shown in Figure \ref{fig:deVriesdiagram}. The average width is $\sim$0.28 $\upmu$m and the average peak position is $\sim$69.2 $\upmu$m. Over-plotted are laboratory measurements from \citet{2006Suto} with an iron fraction of 0\% (right solid line). The right dashed line gives the widths and peak positions of 1\% iron fraction; these are interpolated between 0\% and 8\% iron fractions using a linear trend fitted \citep{2012deVries} to the measured values between 0\% and 16\% iron fractions from \citet{2003Koike}. The 69 $\upmu$m feature shapes of protoplanetary disks are similar to laboratory measurements of $\sim 150-200$ K forsterite with an iron fraction between $0-1\%$. 

The $I_{23}/I_{69}$ ratio and the 69 $\upmu$m band shapes are not consistent for HD\,100546, HD\,144668, HD\,141569 and Oph IRS 48.  Thus, a single-temperature analysis cannot explain the difference in the observational appearance of forsterite, and more complex models have to be invoked. We show in Sects. \ref{sec:modeling_approach} and \ref{sec:relativestrengths} that the $I_{23}/I_{69}$ ratio and the 69 $\upmu$m band shape is sensitive to (distributions) of temperatures, iron fractions, grain sizes, and optical-depth effects. For HD\,100546, it has been shown that the difference in temperature indicators can be explained by optical-depth effects that increases the strength of the 69 $\upmu$m feature compared with the 23 $\upmu$m feature because the disk is optically thinner at longer wavelengths \citep{2011Mulders}. Possibly, a similar scenario may be the case for HD\,144668. However, for the evolved transitional disks HD\,141569 and Oph IRS 48, such optical-depth effects are not sufficient since these disks are much more optically thin. Their very weak short-wavelength features suggest that the forsterite must be cold (T $\lesssim90$ K). However, the band shape of their 69 $\upmu$m features suggests much higher temperatures (174 K and 156 K, \citealt{2013Sturm}). In Sect. \ref{sec:hd141569}, we perform a case study of HD\,141569 and present forsterite models that can fit the short- and long-wavelength spectra.

\begin{table}

\centering
\begin{tabular}{llcccc}
\hline\hline
ID & Object & p$_L$ ($\mu$m) & w$_L$ ($\mu$m) & p$_G$ ($\mu$m) & w$_G$ ($\mu$m) \\
\hline
1 & Oph IRS 48& 69.170& 0.269& 69.168& 0.521\\
2 & HD141569& 69.300& 0.288& 69.303& 0.600\\
3 & HD100546& 69.191& 0.317& 69.194& 0.681\\
4 & HD179218& 69.199& 0.262& 69.196& 0.502\\
5 & HD104237& 69.240& 0.212& 69.224& 0.351\\
6 & HD144668& 69.079& 0.312& 69.088& 0.599\\
\hline
\end{tabular}
\caption{\label{tab:featureshapes}Peak positions and widths given for Lorentzian and Gaussian fits. For clarity, they are both given so that they can be compared with the Lorentzian values derived for Beta Pic \citep{2012deVries}.  Gaussian values are taken from \citet{2013Sturm}. The difference between Lorentzian and Gaussian widths stems from their definition.}

\end{table}

\begin{table*}[htdp]
\tiny
\begin{center}
\begin{tabular}{llccccccccccc }

\hline 
\hline

ID & Target  &  Spitzer &69 $\upmu$m      & $F_{30}/F_{13.5}$  & group &$I_{23}$  &$I_{69}$      &       $L_{*}$& $T_{*}$ & $d$ &        $F_{1300}$    \\ 

\#&             & detections             & detection     &  & &          erg cm$^{-2}$ s$^{-1}$             &     erg cm$^{-2}$ s$^{-1}$                      & L$_{\odot}$ & K & pc  & mJy  \\ 
&               &                               &                        &       $\triangledown$ &        &  &         &                                                       &          &       &    \\  
\hline
1& Oph IRS 48   &       \VVV                    & \VVV &       10.40 $\pm$     0.30          &Ib &$1.14 \pm 0.21 \times 10^{-11}$  &$6.11 \pm 0.21 \times 10^{-13 } $     &         14.3          &       10000   &       120              &  60    $\pm$ 10       \\
2& HD\,141569   &       \VVV                    & \VVV &       6.80 $\pm$      0.20         &Ib &$ 7.67\pm1.53 \times 10^{-12}$           &$6.32 \pm 0.49 \times 10^{-13     }$      &       18.3            &       9520    &       99                &   5 $\pm$ 1 \\
3& HD\,100546   &       \VVV                    &\VVV   &       3.50 $\pm$      0.20         &Ia & $7.08 \pm1.40 \times 10^{-10}$  &$9.46 \pm 0.13 \times 10^{-12 }  $    &         32.9          &       10500   &       103              &   465 $\pm$ 20         \\
4& HD\,179218   &       \VVV                    &\VVV   &       2.40 $\pm$      0.20         &Ia & $ 8.97\pm1.79\times 10^{-11}$   &$7.33 \pm 0.18 \times 10^{-13 }  $    &              79.2             &       9810    &       244         &   71 $\pm$   7         \\ 
5& HD\,104237   &       \VVV                    &\VVV   &       1.28 $\pm$     0.03          &IIa & $4.41\pm 0.88 \times 10^{-11}$         &$2.14 \pm 0.21 \times 10^{-13     }    $  &    34.7               &       8405    &       116       &   92 $\pm$ 19  \\
6& HD\,144668   &       \VVV                    & \VVV  &      0.96 $\pm$     0.02          &IIa & $1.93\pm0.38 \times 10^{-11}$  &$2.60 \pm 0.13 \times 10^{-13 }   $   &       87.5            &       7930    &       208              &      20 $\pm$ 16   \\
\hline
7& HD\,169142   &       \VVV                    &\dots   &       7.80 $\pm$      0.50         &Ib &$1.68\pm 0.34 \times 10^{-11}$    &$\uparrow1.73 \pm 0.56 \times 10^{-13      }  $   &     15.3                      &       8200    &       145       &   197 $\pm$   15      \\
8& HD\,100453   &       \VVV                    & \dots &       5.20 $\pm$      0.30         &Ib &$1.77\pm 0.35 \times 10^{-11}$    &$ \uparrow2.93 \pm 0.34 \times 10^{-13}        $       &         8.0           &       7390    &       112       &    200 $\pm$ 31  \\
9& HD\,142527   &       \VVV                    &\dots   &       5.00 $\pm$      0.10         &Ia &$4.64\pm 0.92 \times 10^{-11}$    &$\uparrow2.67 \pm 0.86 \times 10^{-13     } $     &       50.6            &       6260    &       198       &   1190 $\pm$ 33 \\
10& HD\,36112   &       \VVV                    &\dots   &       4.10 $\pm$      0.20         &Ia &$ 1.40\pm 0.28\times 10^{-11}$    &$\uparrow7.20 \pm 2.30 \times 10^{-14     }   $   &       22.2            &       7850    &       205       &   72 $\pm$   13\\
11& HD\,203024  &       \VVV            &\dots   &       2.30 $\pm$      0.20      &Ia &$ 3.08 \pm 0.62 \times 10^{-11}$          &$\uparrow3.26 \pm 0.41 \times 10^{-13     }  $    &            99.2               &       8200    &       620       &      \dots    \\
12& HD\,163296  &       \VVV            &\dots   &       2.00 $\pm$      0.10      &IIa & $4.26 \pm 0.85 \times 10^{-11}$        &$\uparrow8.20 \pm 2.60 \times 10^{-14}  $         &           23.3                &       8720    &       122       &  743 $\pm$   15 \\
13& HD\,144432  &       \VVV            &\dots   &       1.82 $\pm$     0.06           &IIa & $ 9.36 \pm 1.90\times 10^{-12}$        &$\uparrow8.40 \pm 2.70 \times 10^{-14}    $       &       10.1            &       7345    &       145       &  44 $\pm$   10         \\
14& HD\,142666  &       \VVV            &\dots   &       1.53 $\pm$     0.05           &IIa &$\uparrow 1.73 \pm 0.35\times 10^{-12}$&$\uparrow6.53 \pm 2.11 \times 10^{-13} $   &      14.4            &       7580    &       145         &   127 $\pm$ 9 \\
15& HD\,150193  &       \VVV            &\dots   &       1.42 $\pm$     0.05           &IIa &$ 2.50 \pm 0.50\times 10^{-11}$          &$\uparrow6.70 \pm 2.20 \times 10^{-14      }   $  &    24.0               &       8990    &       150       &   45 $\pm$ 12 \\
16& HD\,50138   &       \VVV                    &\dots   &        0.78 $\pm$     0.04          &IIa &$8.38\pm1.67 \times 10^{-11}$    &$\uparrow 2.08 \pm 0.67 \times 10^{-13}         $      &           423.5       &       12230   &       289       &      \dots             \\
17& HD\,98922   &       \VVV                    &\dots   &      0.75 $\pm$     0.03          &IIa &$3.01\pm 0.25 \times 10^{-11}$   &$\uparrow1.15 \pm 0.37 \times 10^{-13     }    $  &      855.7            &       10500   &       538       &      \dots             \\
\hline
18& HD\,135344\,B       &       \dots           & \dots &       10.90 $\pm$      0.30         &Ib &$\uparrow 3.72 \pm 0.74 \times 10^{-13}$ &$\uparrow1.12 \pm 0.36 \times 10^{-13}$       &     8.3               &       6590    &       140       &   142 $\pm$ 19 \\
19& HD\,97048   &       \dots                   & \dots &       5.90 $\pm$      0.40         &Ib &$\uparrow 9.80\pm 1.90\times 10^{-12}$ &$\uparrow1.96 \pm 0.63 \times 10^{-13 }$ &     40.7           &       10010   &       158         &  452 $\pm$ 34 \\
20& AB Aur              &       \dots   & \dots &       4.50 $\pm$      0.10      &Ia &$ \uparrow 2.19\pm 0.43\times 10^{-11}$ & $\uparrow1.01 \pm 0.09 \times 10^{-12}$       &       46.2            &       9520    &       144              &   136  $\pm$ 15       \\
21& HD\,139614  &       \dots           &\dots  &       4.20 $\pm$      0.30      &Ia &$\uparrow 2.18 \pm 0.43 \times 10^{-12}$& $\uparrow7.50 \pm 2.40 \times 10^{-14}$         &      8.6            &       7850    &       140              &   242 $\pm$ 15   \\
22& HD\,38120   &       \dots                   &\dots  &       2.60 $\pm$      0.20         &Ia &$\uparrow 4.03\pm0.81\times 10^{-12}$     & $\uparrow6.00 \pm 1.90 \times 10^{-14}$         &    89.3     &       10500   &       510         &      \dots    \\
23& HD\,35187   &       \dots                   &\dots   &       2.10 $\pm$      0.10         &IIa& $\uparrow 8.35\pm1.60 \times 10^{-13}$ &$\uparrow1.98 \pm 0.23 \times 10^{-13}$       &    27.4               &       8970    &       150       &   20 $\pm$ 2  \\
\hline

\end{tabular}
\end{center}
\caption{\label{tab:sample} Sample of Herbig stars. $\uparrow$ indicates upper limits. References of mm photometry are given in Table 1 of \citealt{2014Maaskant}.}
\end{table*}


\section{Radiative transfer and dust model}
\label{sec:modeling_approach}
Many physical assumptions are made in models of dust features from protoplanetary disks. In this section we explain our modelling approach. First we discuss the dust model used to analyse forsterite features. Thereafter we briefly explain how forsterite is implemented in the radiative transfer code MCMax \citep{2009Min} .

\subsection{Dust model}
\label{sec:dustmodel}
Throughout this paper, we make use of a standard dust composition that is responsible for the continuum radiation of 80\% silicate and 20\% amorphous carbon. This is the dust component without forsterite. To be consistent with the previous radiative  transfer modelling by \citet{2011Mulders}, the adopted amorphous dust composition with reference to the optical constants is 32\% MgSiO$_3$ \citep{1995Dorschner}, 34\% Mg$_2$SiO$_4$ \citep{1996HenningStognienko}, 12\% MgFeSiO$_4$ \citep{1995Dorschner}, 2\% NaAlSi$_2$O$_6$ \citep{1998Mutschke}, 20\% C \citep{1993Preibisch}. The shape of our particles is irregular and approximated using a distribution of hollow spheres (DHS, \citealt{2005Min}) using a maximum vacuum fraction of f$_{\rm max} = 0.7$. In DHS, f$_{\rm max}$ is the parameter that controls the shape of the particle and can be regarded as the `irregularity' parameter because there is no observable difference between porosity and irregularity.  

\begin{figure}[t] 
\includegraphics[width=\columnwidth]{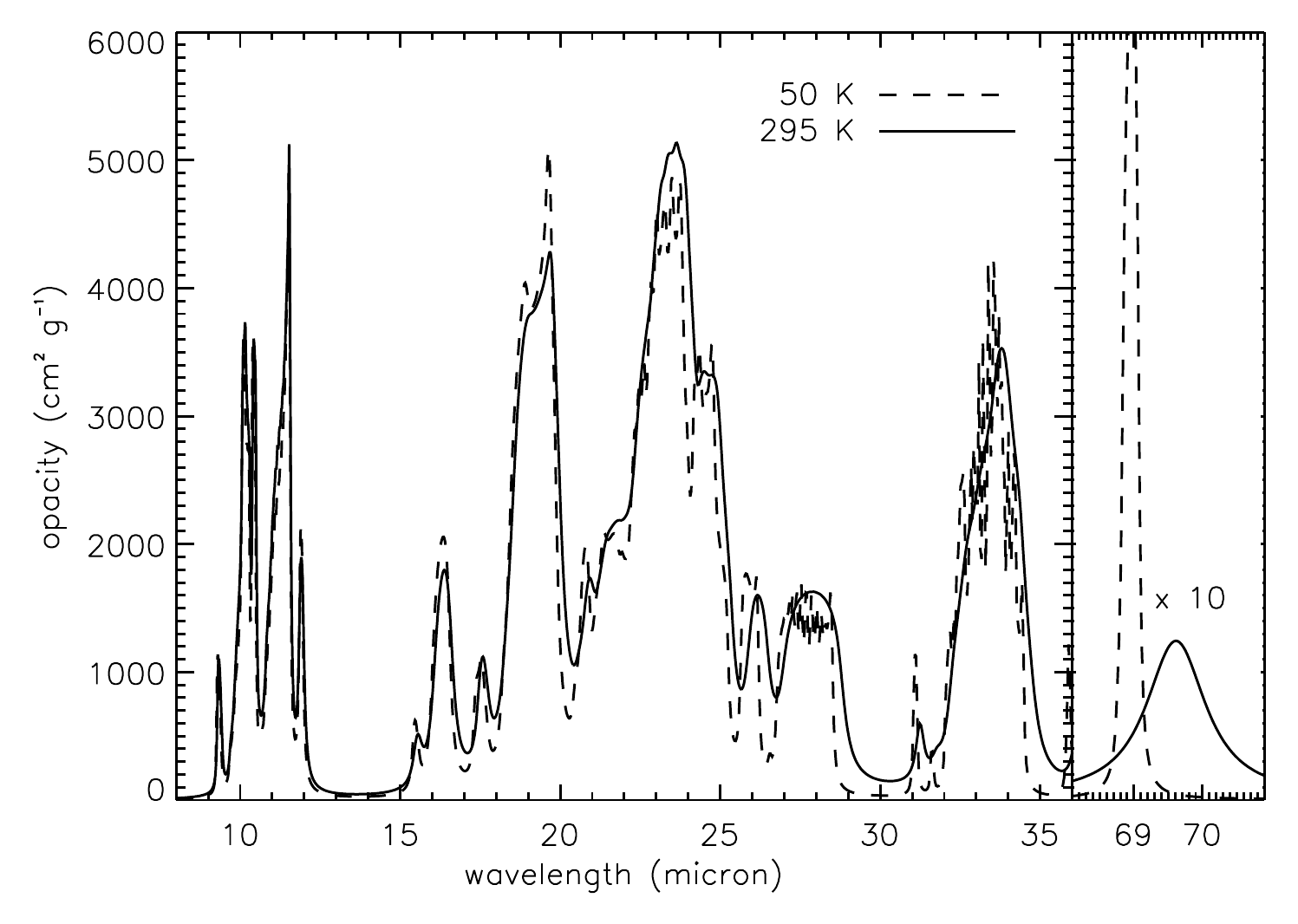}
\includegraphics[width=\columnwidth]{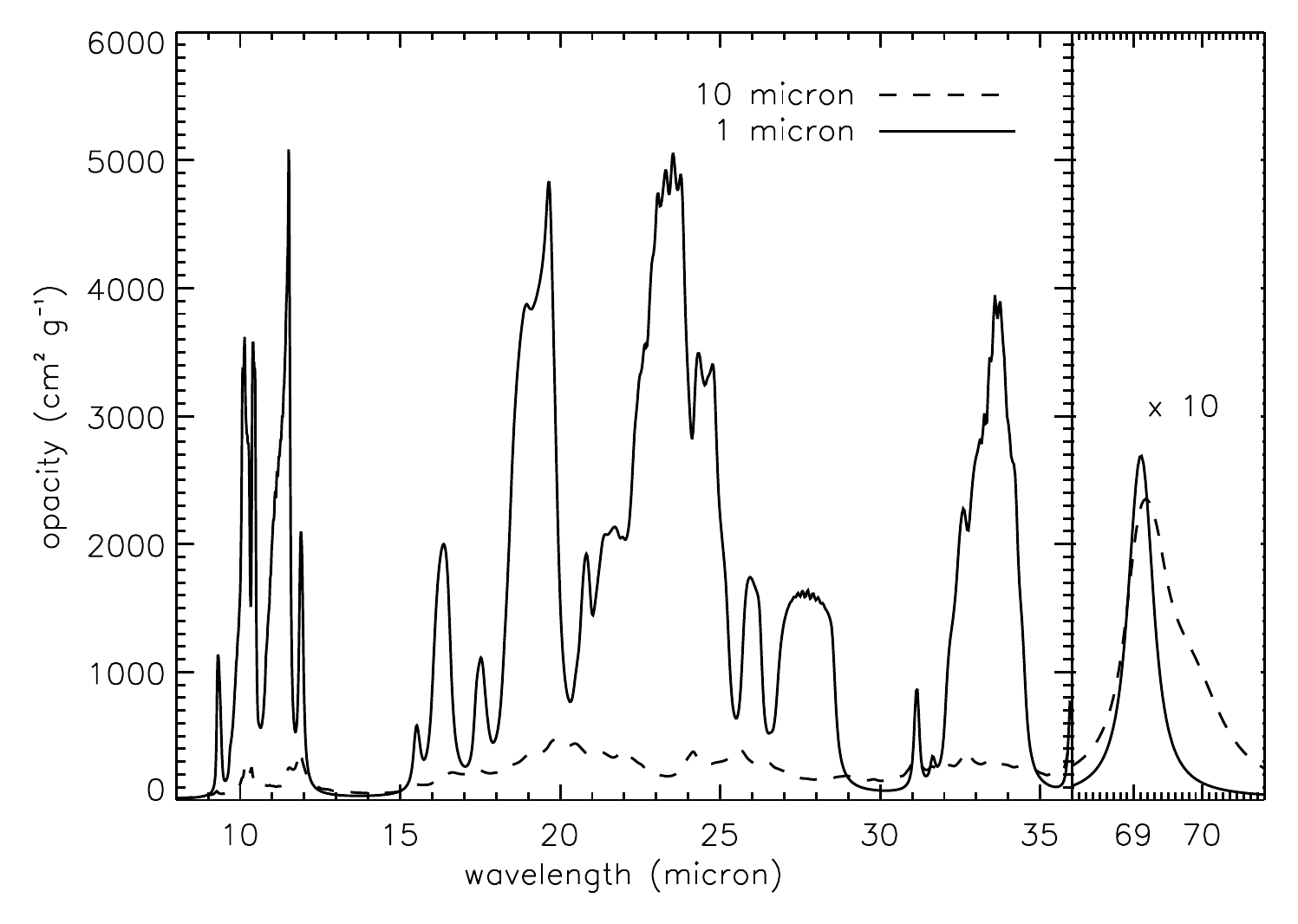}
\caption{\label{fig:opacity} Top: calculated opacities of 1 $\upmu$m forsterite grains at 50 K and 295 K. Bottom: calculated opacities for 1 and 10 $\upmu$m forsterite grains (T = 150 K).} 
\end{figure}

Forsterite features were analysed using the optical constants from laboratory data of \citet{2006Suto} and irregular DHS particles with a maximum vacuum fraction of f$_{\rm max} = 0.7$, which has been found to be a good representation of observed short-wavelength crystalline silicate profiles in Spitzer data \citep{2010Juhasz}. The effects of temperature on the width and peak position of the 69 $\upmu$m feature are also illustrated in Fig. \ref{fig:deVriesdiagram}. The forsterite opacities of \citet{2006Suto} are temperature dependent and are the best opacities available to use in radiative transfer modelling of protoplanetary disks where the temperatures are below $\lesssim 1500 $ K. Figure \ref{fig:opacity} shows the opacities of several forsterite grains. The top figure shows 1 $\upmu$m grains for 50 K and 295 K. The short-wavelength features are very similar, while the peak strength, peak position, and width of the 69 $\upmu$m feature are very sensitive to the temperature in this range. The integrated feature strength of the 69 $\upmu$m feature is $\sim$3 times higher for 50 K than ofr 295 K forsterite. The bottom figure shows the opacities of 1 and 10 $\upmu$m grains. Because of the larger grain size, the grain model predicts that the shorter wavelength features become weaker and that the 69 $\upmu$m feature broadens. A more detailed discussion on the emission characteristics of larger ($\sim$ 10 $\upmu$m) forsterite grains is given in \citet{2004Min}. The temperatures of amorphous silicates and carbonaceous grains were calculated from radiative equilibrium. It is assumed that the forsterite grains are in thermal contact with the other dust constituents. Where relevant, we evaluate in the next sections whether this assumption is reasonable; see \citet{2011Mulders} for an extended discussion on the assumption of thermal contact between the dust grains in radiative transfer modelling of protoplanetary disks.

\subsection{Radiative transfer model}
We used the Monte Carlo radiative transfer code MCMax to compute full radiative transfer models. With a given set of star and disk input parameters, MCMax can self-consistently compute the temperature and the vertical density structure of highly optically thick disks. From these models it produces a range of observables such as the resulting SED and the forsterite features, which are calculated by integrating the formal solution to the equation of radiative transfer by ray-tracing.

\label{sec:diskmodel}


\section{Benchmark modelling}

\label{sec:relativestrengths}
In this section, we construct a benchmark model for a typical disk around a Herbig star. This model is used to investigate the key parameters that change the feature strength ratio of forsterite in protoplanetary disks. The aim is to provide a qualitative description  for the behaviour of the feature strength ratio $I_{23}/I_{69}$. Thereby, we derive estimates of the minimum and maximum $I_{23}/I_{69}$ ratios and compare them with the observed values in the spectra of the objects in our sample. We start by introducing the properties of a benchmark disk model. We study four evolutionary disk scenarios that control the $I_{23}/I_{69}$ band strength ratio: \textbf{1)} grain growth, \textbf{2)} transitional disk, \textbf{3)} radial mixing, and \textbf{4)} optical depth effects. 

\subsection{Benchmark disk model}
We modelled a hydrostatic flaring disk with stellar properties from HD\,97048 (see parameters in Table \ref{tab:sample}). The parameters of this model are taken from the study of \citet{2014Maaskant} and are summarised here (see Table \ref{tab:benchmarkmodel}). The chosen stellar properties ($L_{*} = 40.7 L_{\odot} $, $T_{*} = 10010$ K) are typical for Herbig Ae/Be stars. By this choice, the influence of the stellar radiation on the thermal structure of the disk can be assumed to represent the objects in our sample. The inner and outer radii of the disk were set at 0.1 and 250 AU. The radial dependence of the surface density is parametrised by a power law with index $-1$. The dust mass is M$_{\rm dust} = 3 \times 10^{-4}$ M$_{\odot}$, where we assumed a gas-to-dust ratio of 100. The dust size ranges from $a_{\rm min}= 0.01$ $ \upmu$m up to $a_{\rm max}=1$ mm and follows a power-law distribution with $a_{\rm pow}=-3.5$. 

The standard benchmark model described above does not yet contain forsterite. The models presented in the following sections are different from the benchmark model because forsterite is added to the disk between given radii. This was done by replacing a mass fraction of 10\% of the standard dust component by forsterite grains. Note that the total dust mass at any radius is therefore conserved and that the forsterite grains used in the models are all one size and iron free. Additionally, in the scenarios \textit{transitional disk} and \textit{optical depth}, the disk parameters $R_{in,disk}$ (inner radius of the disk) and $M_{dust,disk}$ (disk mass in dust) are changed to better mimic the disk evolutionary scenarios. All other parameters are held constant so that the effect of the changed parameter is best visible.

\begin{table}[htdp]
\caption{\label{tab:benchmarkmodel}Characteristics of the benchmark model. }
\begin{center}
\begin{tabular}{l l l c c c c c}
\hline
\hline
\multicolumn{2}{c}{Parameter }                                                  & Unit    &Benchmark                 \\
\hline

Stellar temperature             &               $T_{*}$                         & K                               &10 010                  \\     
Stellar luminosity              &               $L_{*} $                                & L$_{\odot}$             & 40.7                            \\
Stellar radius                  &               $R_{*}$                         &R$_{\odot}$            &       2.12                           \\
Stellar mass                    &               $M_{*}$                         &M$_{\odot}$             &       2.50              \\
Inclination                             &               $i$                                     &$^{\circ}$             &45                     \\
Distance                                &               $d$                                     &pc                                 &158            \\
\vspace{0.001mm}\\
Disk, inner radius      &               $R_{in,disk}$                   &AU                             &0.1                                 \\
Disk, outer radius      &               $R_{out,disk}$                  &AU                             &250                          \\
Silicate fraction                       &               $f_{Si}$                                & \dots           &0.8             \\
Carbon fraction                 &               $f_{C}$                         & \dots           & 0.2              \\
Min dust size                   &                $a_{min}$                      & $\upmu$m        &  0.01                                 \\
Max dust size                   &                $a_{max}$                      &  mm             &   1            \\
Dust-size power-law index&              $a_{pow}$                       &         \dots           & -3.5  \\
Disk, gas mass          &               $M_{gas,disk}$                  & M$_{\odot}$             & $3.0\times10^{-2}$  \\                
Disk, dust mass &               $M_{dust,disk}$         &  M$_{\odot}$          &$3.0\times10^{-4}$  \\

 \hline
\end{tabular}
\end{center}
\end{table}%

\begin{figure*}[t] 
\centering
\includegraphics[width=0.6\textwidth]{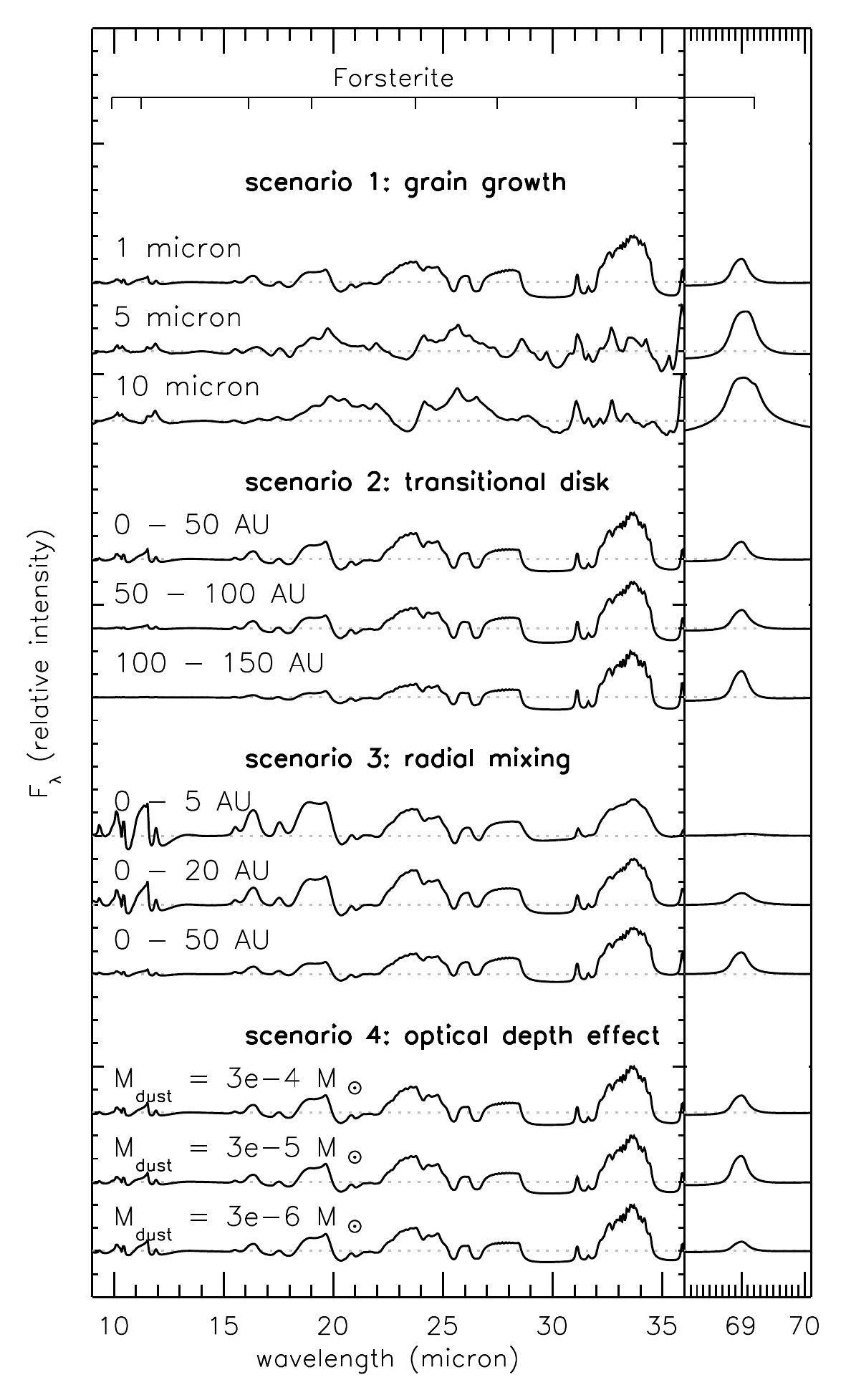}
\hspace{0.05\textwidth}
\includegraphics[width=0.28\textwidth]{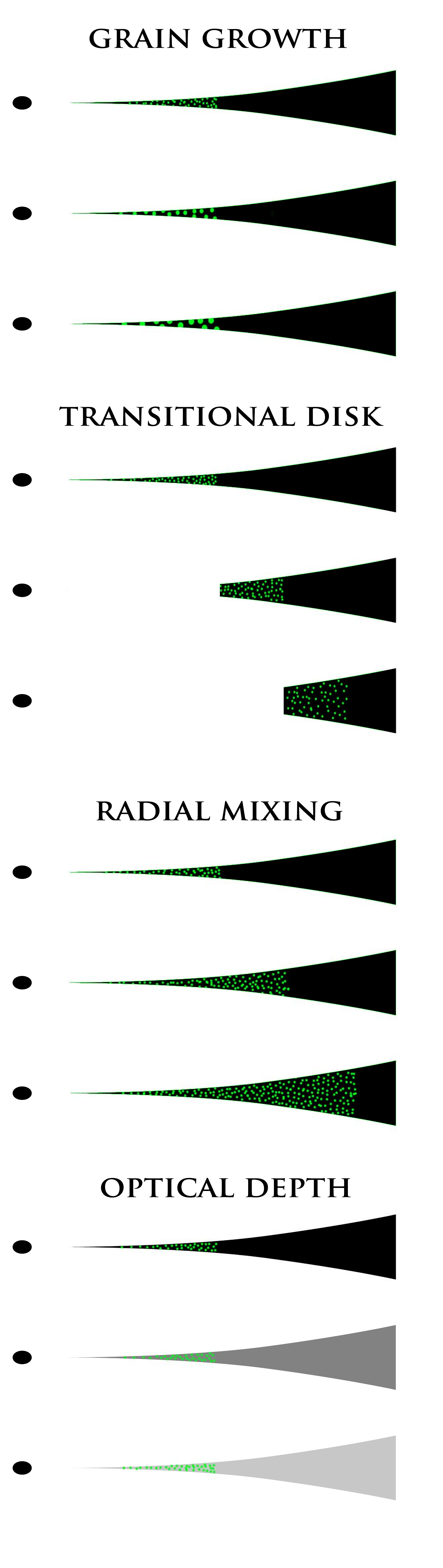}
\caption{\label{fig:scenarios} Grain-size,  transitional-disk,  and radial-mixing scenario; optical-depth effect. The continuum-subtracted spectra are scaled relative to the strongest band so that the effect on the forsterite temperature  is presented best. The right panel shows sketches to illustrate the structure of the disk and the location and sizes of forsterite grains.} 
\end{figure*}

\subsection{Scenario 1: grain growth}
\label{sec:graingrowth}

In the \textit{grain-growth} scenario, the dependence of the relative feature strengths on the grain size is studied. We compared forsterite grains of 1, 5, and 10 $\upmu$m in size.  In this modelling grid, forsterite is mixed in the region between $0-50$ AU. As shown by the opacities in Fig. \ref{fig:opacity}, for grains $\gtrsim 1$~$\upmu$m, the shorter wavelength features are weaker than the 69 $\upmu$m feature. Thus $I_{23}/I_{69}$ also decreases. For grains smaller than a few micron, the relative band strength is constant. Because the features between $25-35$ $\upmu$m change in position, the points where the continuum is subtracted is changed from 21.75 and 26.4 $\upmu$m to 23.5 and 29 $\upmu$m.

\subsection{Scenario 2:  transitional disk}
The \textit{transitional-disk} scenario highlights the temperature dependence of the feature strength ratios. To model this, we created disk models with increasingly larger inner holes (50, 100, and 150 AU). We then assumed in each transitional disk that forsterite is present (10 \%) in the 50 AU-wide region just outside of the gap. For the disk as a whole this means that an increasing amount of near-infrared continuum is missing. The dominant effect for the forsterite feature ratios is the temperature selection that is taking place.

\subsection{Scenario 3: radial mixing}
\label{sec:radialmixing}
The effect that forsterite is gradually mixed towards larger radii is studied in the \textit{radial-mixing} scenario. To model this scenario, a forsterite mass abundance of 10\% wa mixed in the disk between 0.1 AU up to 5, 20 and 50 AU.  When forsterite is mixed with a constant mass abundance ratio up to 50 AU in the disk, the feature strength ratio is $I_{23}/I_{69}= 1.2$. If forsterite is only present in the inner disk within 5 AU, the 69 $\upmu$m feature is relatively weak and $I_{23}/I_{69}= 15.2$.

\subsection{Scenario 4: optical-depth effects}
\label{sec:opt_depth}
The \textit{optical-depth-effects} scenario investigates the feature strength ratio as a result of the differences in dust-continuum optical depth. The optical depth is lower at longer wavelengths, therefore the $\tau =1$ surface at 69 $\upmu$m lies deeper in the disk than the $\tau =1$ surface at 23 $\upmu$m. Hence, the regions in the disk that produce the 23 and 69 $\upmu$m features have different volumes. Thus the feature strength ratio can be altered if the optical depth changes in the disk (as shown for HD\,100546 in \citealt{2011Mulders}). We modelled this scenario by adopting a mass abundance of 10\% of forsterite within $0-50$ AU in the disk, and by varying the mass of the disk between 3$ \times 10^{-4}$ M$_{\odot}$ and 3 $\times 10^{-6}$ M$_{\odot}$. By decreasing the total mass in the disk, the disk moves stepwise from optically thick at both 23 and 69 $\upmu$m to optically thin at 69 $\upmu$m, and on to optically thin in both features. The 1.3 mm luminosity values of these disk models are over-plotted in Fig. \ref{fig:L_vs_ratio1} and cover the range of observed 1.3 mm continuum detections for our sample of Herbig stars.

\subsection{Summary of modelling scenarios}
Comparing the results in this section with the observations as presented in Table \ref{tab:sample} and . \ref{fig:69_vs_forstT} and \ref{fig:L_vs_ratio1}, we obtain the following results:

Since grains grow in protoplanetary disks, the \textit{grain-growth} scenario may be important to explain the observed forsterite feature strength ratios. Figure \ref{fig:scenarios} shows the relative band strengths for different grain sizes. We find that for 1 $\upmu$m grains the $I_{23}/I_{69}$ band strength ratio is highest.

For grains larger than a few micron, we find that the 23 $\upmu$m band disappears and thus that the $I_{23}/I_{69}$ ratio decreases to zero. For one particular object, HD\,141569, it is demonstrated in Sect. \ref{sec:hd141569} that the grain size is a plausible explanation for the feature strength ratio. Dust-settling and grain growth of the standard dust component are not included. In reality, other grains also grow in size, and larger grains settle towards the midplane, leaving relatively more small dust grains in the surface layers of the disk. These effects mainly affect the optical depth of the disk. Therefore, grain growth and settling change the ratio of flux contributions from small and larger grains. Thus the final spectrum may be expected to be a balance between small and large grains, set by the specific disk parameters.  

In the \textit{transitional-disk} scenario, dust is cleared from the inner region and the forsterite is gradually mixed at larger radii. Therefore these models show the effect of a decreasing average forsterite temperature.  This scenario is limited in understanding the origin of the gap and the forsterite in the outer disk. It is yet not understood, for example, if the presence of cold forsterite is due to in situ formation in the outer disk or is radially mixed outward. Figure \ref{fig:scenarios} shows that if the forsterite occurs at larger radii in the disk, the $I_{23}/I_{69}$ ratio decreases. Since the temperature is lower in the outer disk, the peak of the disk emission shifts towards longer wavelengths and the strength of the 69 $\upmu$m feature increases in strength compared with that of the 23 $\upmu$m feature. This may partly explain the forsterite spectra of transitional disks with known large gaps such as Oph IRS 48 and HD\,141569.

An important insight follows from the  \textit{radial-mixing} modelling grid for objects with strong forsterite features detected in the Spitzer/IRS spectrum, but no 69 $\upmu$m feature detections. Figure \ref{fig:scenarios} shows that if forsterite is only present in the inner disk, the 69 $\upmu$m feature is much weaker than the 23 $\upmu$m feature. There are two reasons for this effect. The first is based on the temperature. Even if the forsterite is hot, the 69 $\upmu$m feature may not be visible in the spectrum because continuum emission from colder regions of the disk (without forsterite) dominates. The second effect is that the integrated strength of the 69 $\upmu$m band is three times weaker for 295 K than for 50 K forsterite (see Sect. \ref{sec:dustmodel}). In addition, because the feature broadens towards higher temperatures, the peak flux of the 69 $\upmu$m feature decreases by almost an order of magnitude (see Fig. \ref{fig:opacity}). It is thus expected that the 69 $\upmu$m feature will further broaden and decrease in strength for temperatures above 295 K. The low upper limits of the 69 $\upmu$m detections typically show very high values of $I_{23}/I_{69} \sim 100$ (most notably for Group II sources). Since a forsterite abundance of 10 \% between 0--5 AU gives a $I_{23}/I_{69}= 15.2$, this suggests that forsterite must be even closer to the star ($\sim1-2$ AU). Because forsterite only resides in these hot inner regions of the disk, radial mixing is likely not an efficient process. 

We find that within a limited mass range, \textit{optical-depth} effects can enhance the 69 $\upmu$m feature as it becomes optically thin while the disk is still optically thick at 23 $\upmu$m. This may explain the higher detection rate of the 69 $\upmu$m feature for disks with a lower mm luminosity (Fig. \ref{fig:L_vs_ratio1}). There are three principal cases for the relative optical depth at 23 and 69 $\upmu$m. The first situation is that the disk is optically thick at both 23 and 69 $\upmu$m. This is the case for a disk mass of M$_{\rm dust} = 3 \times 10^{-4}$ M$_{\odot}$. The optical depths scale relative to the density. Assuming that forsterite is vertically well mixed, the feature strength ratio is quite constant. The second situation is that the disk becomes optically thin at 69 $\upmu$m while it is still optically thick at 23 $\upmu$m (M$_{\rm dust} = 3 \times 10^{-5}$ M$_{\odot}$). In this case,  the 69 $\upmu$m feature becomes stronger and the feature strength ratio decreases. Third, the disk mass is so low that the disk is optically thin at both wavelengths. In this situation, all forsterite is visible to the observer and the feature strength ratio increases again.

Models predicting strong 69 $\upmu$m features and weak features at shorter wavelengths are the grain-growth scenario (where the emission is dominated by larger $\gtrsim$ 1 $\upmu$m forsterite dust),  the transitional disk scenario (where only cold forsterite is left in the disk), and the optical-depth scenario (the mass of small dust grains decreases and the disk becomes optically thin at 69 $\upmu$m while it is still optically thick at 23 $\upmu$m). 

The opposite case, where the 69 $\upmu$m feature is weaker than
the strong short-wavelength forsterite bands can be explained by inefficient radial mixing. In this scenario, forsterite is only present in the inner disk within a few AU. The location in the disk where the forsterite is hot depends on the stellar properties and geometry of the inner disk. For the continuous-disk models presented in this section, the surface temperature of the disk is $\gtrsim 200$ K at radii $\lesssim 3$ AU. If the grains in the disk are mixed outward, then the 69 $\upmu$m feature rapidly increases in strength. A mixing of 10 \% forsterite in the inner 5 AU would result in a $I_{23}/I_{69}$ ratio of 15.2. Since flat disks have ratios that are an order of magnitude higher, $I_{23}/I_{69}$  (Fig. \ref{fig:69_vs_forstT}), we conclude that mixing is inefficient in flat disks.


\section{Strong and broad 69 micron band of HD141569}
\label{sec:hd141569}
In this section, detailed radiative transfer models of HD141569 are presented. HD141569 was selected for a case study because we found in Sect. \ref{sec:observations} that it is the most extreme outlier in the comparison between the forsterite temperature derived from the $I_{23}/I_{69}$ feature strength ratio (T $\lesssim 70$ K) and the temperature estimate obtained from the shape of the 69 $\upmu$m feature (T $\sim174$ K).

\begin{figure}[t] 
\centering
\includegraphics[trim=0.5cm 0.2cm 0.9cm 0.5cm, width=0.9\columnwidth]{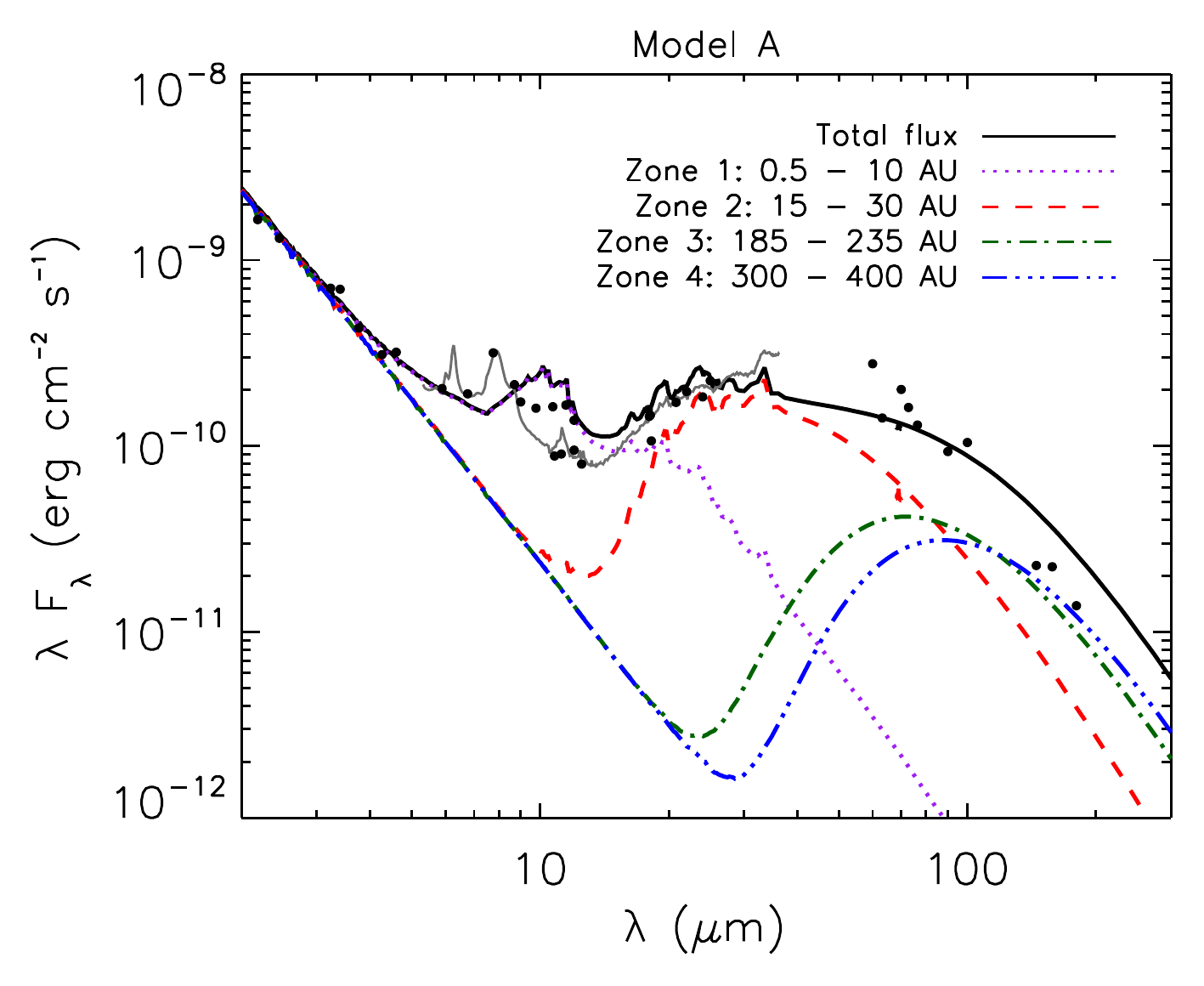}
\includegraphics[trim=0.5cm 0.2cm 0.9cm 0.5cm, width=0.9\columnwidth]{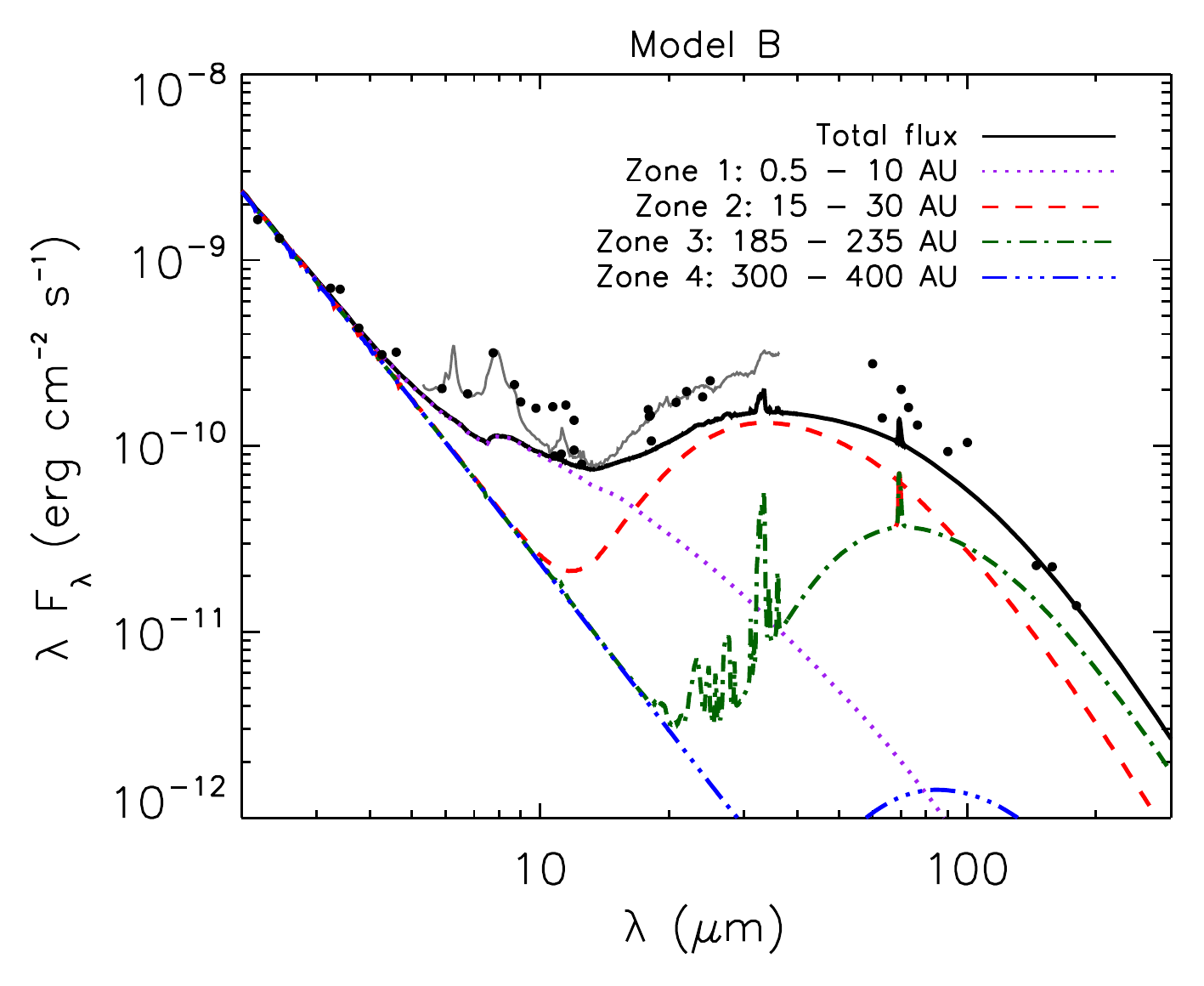}
\includegraphics[trim=0.5cm 0.2cm 0.9cm 0.5cm, width=0.9\columnwidth]{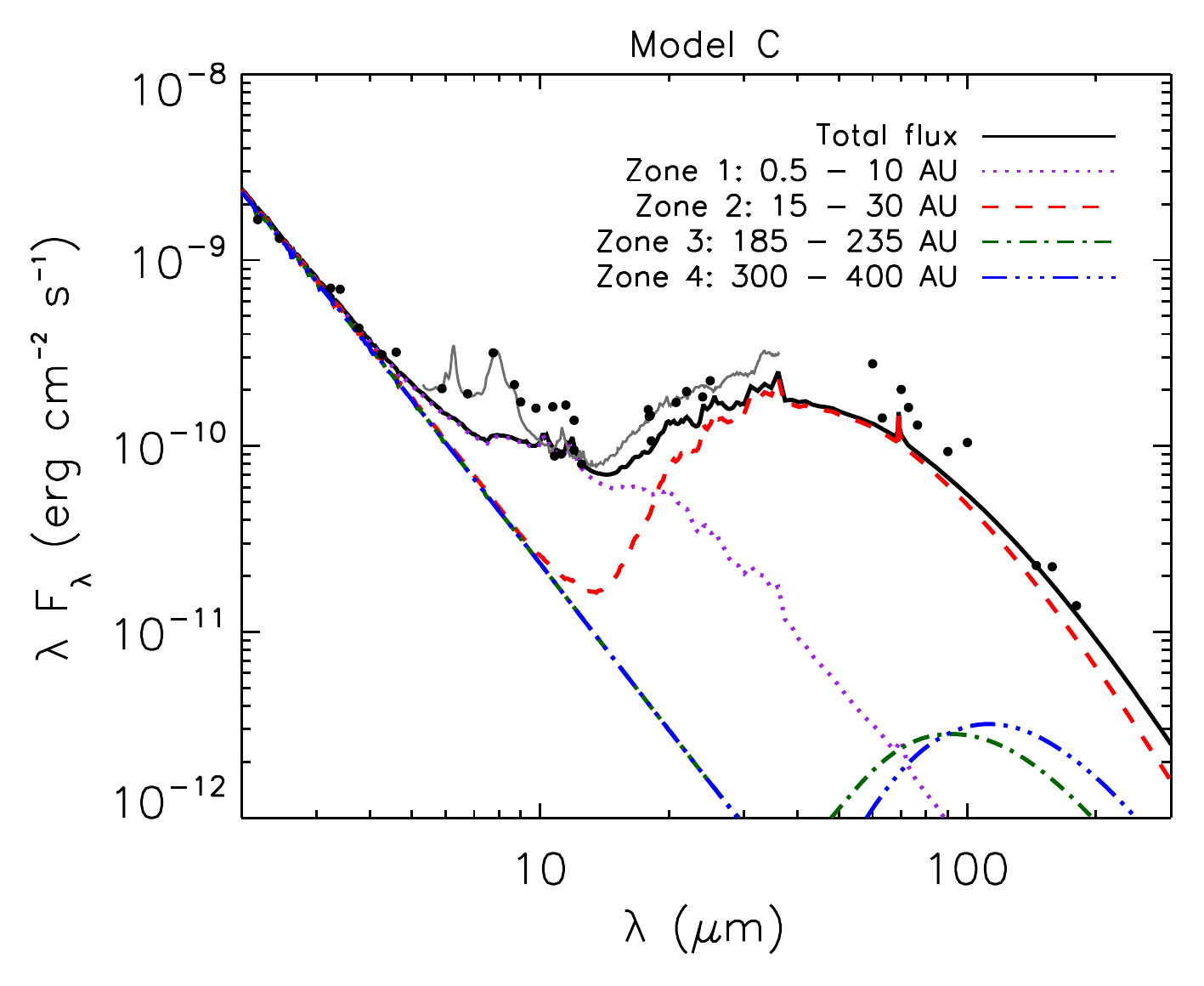}
\caption{\label{fig:SED_hd141569}SEDs of three models of HD141569. Top: model A, small (1~$\upmu$m) forsterite grains. Middle: model B, small (1~$\upmu$m), cold, iron rich forsterite grains. Bottom: model C, large (10~$\upmu$m) forsterite grains. }
\end{figure}

\subsection{Disk-modelling approach}

We focused on the dust mineralogy to constrain the size and location of forsterite. We first derived the geometry of the disk responsible for the dust continuum emission. Thereafter, we explored the properties of the grains that can fit the forsterite spectrum. We searched for solutions that can reproduce the relative band strength $I_{23}/I_{69}$, the shape of the 69 $\upmu$m feature, and simultaneously fit the SED.
  
From earlier studies of the disk of HD141569, it is known that the disk of HD141569 is gas rich and is transitioning to a debris-disk system (e.g. \citealt{1999Augereau, 2000Fisher, 2003Clampin, 2014Thi}). Since we focused on the properties of the dust, our first step was to construct a geometrical model of the dust disk of HD\,141569. We used the stellar properties derived in \citet{1998Ancker}, with a temperature $T_*=9520$ K, luminosity $L_*=18.3$ L$_{\odot,}$ and distance $d=99$ pc. The radiative transfer and dust model are described in Sect. \ref{sec:modeling_approach}. The modelling approach is largely similar to the benchmark model described in Sect. \ref{sec:relativestrengths}. The radial dependence of the surface density is parametrised by a power law with index $-1$. We assumed a gas-to-dust ratio of 100. The dust size ranges from $a_{\rm min}$ up to $a_{\rm max}$ and follows a power-law distribution with $a_{\rm pow}=-3.5$. The largest grain size $a_{\rm max}$ was set at 1 mm. The smallest grain size of the amorphous dust, as well as the forsterite grains, was determined through simultaneously fitting the forsterite features and the SED. Note that the smallest sizes of the forsterite and the amorphous grains were kept the same. The disk mass was also derived by SED-fitting. We assumed an amorphous dust composition of 80\% silicate and 20\% carbon for the dust that emits the continuum emission. Forsterite grains with either a$_{\rm min}$ of 1 $\upmu$m or 10 $\upmu$m were mixed with the disk at certain radii to fit the forsterite spectrum.

\subsection{Disk geometry}

We constructed a disk model that was based on available imaging data and the SED.  To successfully fit the continuum emission of the SED, we found that it is required to break up the disk into four zones, with the following characteristics: zone 1, between $0.5-10$ AU, is the innermost disk and is derived by a fit to the SED between $\sim5-12$ $\upmu$m. Therefore, zone 1 has a typical dust temperature between $\sim150-1000$ K. Note that the continuum emission between the PAH features does not originate from the PAHs \citep{2013Maaskant}. We did not fit the PAH features in this model. Disk zone 2, between $15-30$ AU, accounts for the MIR emission bump between $\sim 12$ $\upmu$m up to at least $\sim100$ $\upmu$m. This disk zone is inferred from the bump in the SED, which is indicative of a high vertical disk-wall that produces a strong single-temperature emission component at $\sim$100 K. Mid-infrared imaging observations by \citet{2000Fisher} are consistent with the presence of such a disk zone, since quadratic deconvolution of the source with the PSF gives FWHM sizes of 34 AU at 10 $\upmu$m and 62 AU at 18 $\upmu$m for the extended emission near the star. However, we did not model the MIR observations, so we cannot accurately determine the inner and outer location of disk zone 2. Therefore we arbitrarily set zone 2 to be located between $15-30$ AU, which results in typical dust temperatures of $\sim70-100$ K. Disk zones 3 and zone 4, located at $185-235$ AU and $300-400$ AU, were inferred from direct scattered-light observations \citep{1999Weinberger, 1999Augereau, 2003Clampin}. From our model, we derive typical temperatures in zone 3 and 4 of $\sim$ 40 K and $\sim$30 K.

Figure \ref{fig:SED_hd141569} shows the three SEDs of forsterite models we present in the next section. A degeneracy that follows from the SED fit is whether the outermost zones 3 and 4 dominate the continuum at long wavelengths. The observed photometry can be fitted by the Rayleigh-Jeans tails of disk zones 2, 3, and 4, as shown in Figure \ref{fig:SED_hd141569}. Millimetre imaging is needed to lift this degeneracy.

\subsection{Fitting the forsterite}

We reproduced the low forsterite $I_{23}/I_{69}$ band strength ratio as well as the shape (i.e. the width and peak position) of the 69 $\upmu$m feature. We did not aim to derive a perfect fit. Instead, we identified the most important parameters that play a role in fitting the forsterite spectrum. Referring back to the parameter study presented in the previous section, we can already indicate in which regions of the parameter space we can find solutions. In general, a low $I_{23}/I_{69}$ ratio can be explained by cold forsterite (as in the transitional disk scenario), or larger grains. Optical-depth effects do not play a role because we found in our models that because of the low mm luminosity, the disk mass is also very low and therefore the disk is optically thin. 

There are only two locations in the disk from which forsterite emission can observed in the spectrum: from disk zone 2, which dominates the SED at MIR wavelengths, and from disk zone 3, which contributes less to the SED because the dust is colder. Forsterite emission from disk zones 1 and 4 is too hot and cold to be observed.

To be consistent with the shape of the forsterite 69 $\upmu$m feature, we needed to include an effect that broadens the band while the forsterite is cold to ensure a low $I_{23}/I_{69}$ ratio. There are two effects that can produce this result. The first solution is to include an iron distribution. This has been described in detail by \citet{2013Sturm}, who found that an iron fraction of between 0--1.2 \% (derived by interpolation between laboratory measurements between 0--8 \%) of forsterite can fit the 69 $\upmu$m feature of HD\,141569.  The second solution is to include forsterite grain size larger than a few micron (Section \ref{sec:graingrowth}). 

\begin{figure*}[t] 
\centering
\includegraphics[width=0.6\textwidth]{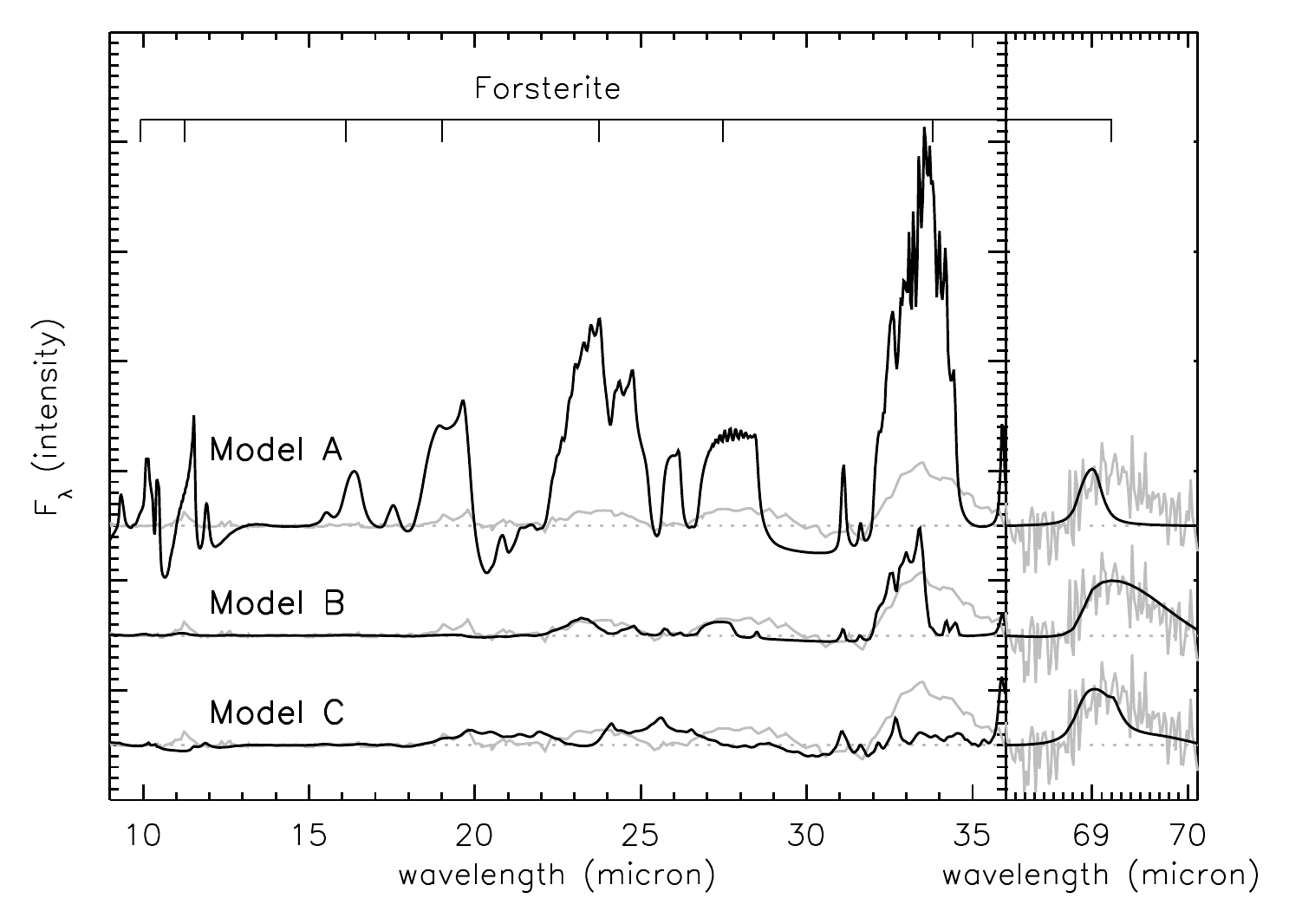}
\includegraphics[width=0.39\textwidth]{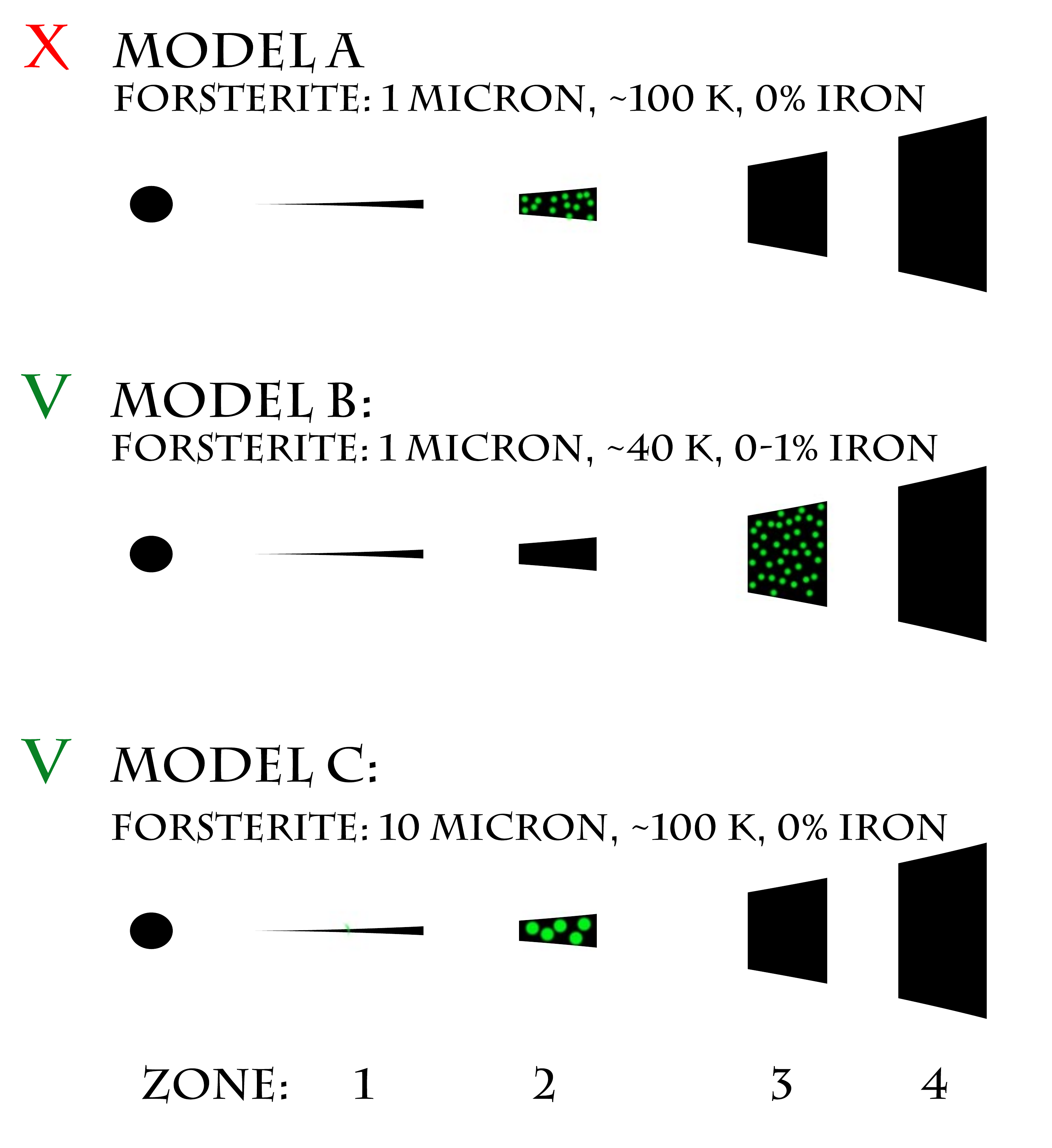}
\caption{\label{fig:HD141569_model_fits} Forsterite spectra of models A, B, and C to HD141569. The black line depicts the model and the grey line the data. Model A fails to fit the data, while model B and C fit the data reasonably well. See text for model details. The right panel shows schematic representations of the disk structures.  Four disk zones are needed to fit the SED and direct-imaging observations. The `X' symbol means that the model fails to fit the spectrum, the `V' symbols shows the best-fitting models. 
 }
\end{figure*}

In the next sections we present three forsterite models. Model A shows that a grain size a$_{\rm min}$ of 1 $\upmu$m forsterite grains in disk zone 2 does not fit the observed spectra. Then we present two different models that can fit the spectrum. Model B, with very cold, 1 $\upmu$m, iron-containing forsterite in disk zone 3, and Model C, with forsterite grains of 10 $\upmu$m mixed in disk zone 2. The parameters of the three models that are used to fit the forsterite spectrum, as well as the SED, are shown in Table \ref{tab:hd141569model}. Spectra and sketches of these models are shown in Figure \ref{fig:HD141569_model_fits}. In the discussion  (Sect. \ref{sec:largergrains}) we evaluate the model solutions, discuss which formation histories are required for both scenarios, and argue that model C, using larger 10 $\upmu$m grains, is favoured.

\begin{table}[t]
\caption{\label{tab:hd141569model}Characteristics of three HD\,141569 models. }
\begin{center}
\begin{tabular}{l c  c c c c c c}
\hline
\hline
\multicolumn{1}{c}{Parameter }  & Unit          &Model A                 &Model B                &Model C                   \\
\hline
Disk total dust mass            &M$_{\odot}$            &3.1    $\times10^{-6}$&1.2$\times10^{-6}$      &9.5$\times10^{-6}$\\
\vspace{0.0001mm}\\
\textbf{Disk zone 1}            &                               &                       &                       &                               \\
Disk location                   &AU                             &0.5 -- 10              &0.5 -- 10           &0.5 -- 10                      \\
Disk, dust mass         &M$_{\odot}$            &5$\times10^{-10}$&7$\times10^{-10}$         &7$\times10^{-10}$   \\ 
Forsterite fraction             & \dots                 &0                      &0                         &0                               \\
Min. dust size                  & $\upmu$m              &  1                    &  10             &  10                   \\
\vspace{0.001mm}\\
\textbf{Disk zone 2}            &                               &                       &                       &                               \\
Disk location                   &AU                             &15 -- 30                 &15 -- 30               &15 -- 30                       \\
Disk, dust mass         & M$_{\odot}$           &1$\times10^{-7}$       &1$\times10^{-7}$       &5$\times10^{-7}$  \\ 
Forsterite fraction             & \dots                 &0.07           &0                         &0.07                            \\
Min. dust size                  & $\upmu$m              &  1                    &  10                     &  10                   \\
\vspace{0.001mm}\\
\textbf{Disk zone 3}            &                               &                       &                       &                               \\
Disk location                   &AU                             &185 -- 235     &185 -- 235  &185 -- 235             \\
Disk, dust mass         &M$_{\odot}$            &1$\times10^{-6}$       &1$\times10^{-6}$       &3.0$\times10^{-6}$  \\ 
Forsterite fraction             & \dots                 &0                      &0.2                         &0                               \\
Min. dust size                  & $\upmu$m              &  1                    &  1                      &  10                   \\
\vspace{0.001mm}\\
\textbf{Disk zone 4}            &                               &                       &                       &                               \\
Disk location                   &AU                             &300 -- 400     &300 -- 400  &300 -- 400             \\
Disk, dust mass         &M$_{\odot}$            &2$\times10^{-6}$       &1$\times10^{-7}$       &6.0$\times10^{-6}$  \\ 
Forsterite fraction             & \dots                 &0                      &0                         &0                               \\
Min. dust size                  & $\upmu$m              &  1                    & 1                       &  10                   \\
 \hline
\end{tabular}
\end{center}
\end{table}%

\subsubsection{Model A: small forsterite grains}
Model A incorporates (iron-free) 1 $\upmu$m forsterite grains in disk zone 2. A mass abundance of 7\% forsterite in zone 2 of the disk was needed to fit the peak flux of the 69 $\upmu$m flux. This model poses several problems. As Figure \ref{fig:HD141569_model_fits} shows, the forsterite features at short wavelengths are over-predicted and the shape of the 69 $\upmu$m band does not match the observation because the forsterite is too warm.  The peak position of the 69 $\upmu$m feature is too blue and the width is too narrow. Because the peak position is sensitive to the iron content of forsterite, this can be alleviated by including a distribution of iron (Sect. \ref{sec:hd141569modelB}).  

The upper plot of Fig. \ref{fig:SED_hd141569} shows that the SED of the disk is only poorly fitted. Because we assumed that the smallest size of the amorphous silicate dust grains is equal to the forsterite grains, the 10 $\upmu$m amorphous silicate features is detected in the spectrum. This feature originates from zone 1, thus we find that the innermost disk zone 1 is depleted of small ($\lesssim$1 $\upmu$m) dust grains. Second, if the outermost zones 3 and 4 contribute $\sim$50 \% of the emission at 69 $\upmu$m,  the SED at mm wavelengths is over-predicted. Thus, we learn that disk zone 2 dominates the continuum emission in the MIR and FIR from $\sim14$ $\upmu$m up to at least $\sim$100 $\upmu$m, and that the contribution from disk zones 3 and 4 at these wavelengths is weak.

The assumption of thermal contact between the forsterite grains and the other disk components has consequences for the emission spectrum of forsterite. For this model, we tested the scenario where the forsterite is not in thermal contact with the amorphous dust. We found that the temperature, and more noticeably, the luminosity of forsterite decreases. Nevertheless, after scaling the abundance to fit the 69 $\upmu$m feature again, the forsterite is still too warm (the short-wavelength features are still over-predicted).

\subsubsection{Model B: small, cold, iron-rich forsterite grains}
\label{sec:hd141569modelB}
In model B, 1 $\upmu$m forsterite grains are mixed in zone 3, where they are farther away from the star and thus colder. To fit the 69 $\upmu$m feature, an abundance of 20 \% forsterite in zone 3 is needed to fit the strength of the 69 $\upmu$m feature. Because of the lower temperature in zone 3, the short-wavelength features are much weaker. To prevent the model from producing the amorphous 10 and 20 $\upmu$m silicate features, the smallest grain sizes in the innermost zones 1 and 2 were set at 10 $\upmu$m. Note that this model implies that there is no contribution from warmer crystalline silicate grains in zone 2.

The shape of the 69 $\upmu$m feature was fitted by assuming that the forsterite has a Gaussian distribution of different iron fractions. It has been shown by laboratory measurements that already an iron fraction of 8 \% shifts the position of the 69 $\upmu$m feature from 69 to 72 $\upmu$m \citep{2003Koike}. Since iron increases the peak wavelength to the red, a Gaussian distribution between 0 and 1 \% iron can broaden the feature significantly. In Fig. \ref{fig:HD141569_model_fits} it is shown that a distribution of iron fractions between $0-1$~\% fits the feature well. The shape of the 69 $\upmu$m was calculated using weighted interpolations of different iron fractions. The interpolation method is described in \citet{2012deVries}.

\subsubsection{Model C: large forsterite grains}
Model C presents a solution using iron-free forsterite grains of 10 $\upmu$m in disk zone 2, using a mass abundance of 7\% to fit the strength of the 69 $\upmu$m feature. Similar to the grain size of forsterite, the smallest grain size of the amorphous dust in the innermost zone 1 and 2 were set at 10 $\upmu$m. For consistency, the smallest grain size in disk zones 3 and 4 were also set at 10 $\upmu$m. The increased grain size influences the opacity profile of forsterite (see Figure \ref{fig:opacity}). The shorter wavelength features are much weaker and the width of the 69 $\upmu$m feature is naturally fitted, although the peak position of the model is slightly too blue ($\sim0.2$ $\upmu$m) compared with the observed feature.  Because the features between $25-35$ $\upmu$m seem to change in position, the points where the continuum was subtracted for this model were changed from 21.75 and 26.4 $\upmu$m to 23.5 and 29 $\upmu$m.

\subsection{Summary of HD 141569 modelling}
Two models fit the observed spectrum of HD\,141569 reasonably
well: model B, which uses iron-rich, $\sim40$ K, 1 $\upmu$m forsterite in zone 3, and model C, which uses iron-free $\sim100$ K, 10 $\upmu$m forsterite in zone 2. Note that any model that combines the parameter sets of these two cases would likely fit the observations
as well. For example, the short-wavelength forsterite features are only poorly fitted in model C, but this problem can be resolved by adding a few small 1 $\upmu$m forsterite grains (while the 69 $\upmu$m feature is still dominated by the larger $\sim$ 10 $\upmu$m forsterite grains). Forsterite formation scenarios will have to explain how forsterite can be present at large distances from the central star and why they are either iron rich, or depleted of small grains. In Sect. \ref{sec:largergrains} we discuss both scenarios. 


\section{Discussion}
\label{sec:discussion}

The composition and sizes of forsterite grains are results of physical processes that take place during the evolution of protoplanetary disks. In the inner regions of disks where temperatures are high ($\gtrsim1000$ K), forsterite can be formed through gas-phase condensation and/or thermal annealing. The crystals may then be transported to the cooler outer parts of the disk. Scenarios include radial mixing in the disk (e.g., \citealt{2002Bockelee-Morvan, 2004Gail, 2006Davoisne, 2014Jacquet}) and disk winds (e.g. \citealt{1994Shu}). Alternatively, crystalline silicates can be formed in situ in the outer disk by shock waves \citep{2002HarkerDesch, 2005Desch}, lightning (e.g. \citealt{1998Pilipp}), parent-body collisions \citep{1967Urey, 2001Huss, 2010Morlok}, and stellar outbursts \citep{2012Juhasz}. To understand the significance of these scenarios, it is important to characterise the location, composition, and sizes of forsterite grains in protoplanetary disks.

\subsection{Forsterite in flat disks}

We found that the $I_{23}/I_{69}$ ratios of flat disks in our sample indicate high ($\gtrsim200$ K) forsterite temperatures. Since such temperatures are only found in the inner $\lesssim5$ AU of a continuous disk (i.e. no gaps), this shows that forsterite emission predominantly originates from the inner region in flat disks. Hence, the observations suggest that for flat disks, forsterite formation occurs in the hot inner region of the disk. 

We found from the radiative transfer models in Sect. \ref{sec:relativestrengths} that the 69 $\upmu$m feature is weakest if forsterite is only present in the inner few AU of the disk. If forsterite becomes abundant in colder regions of the disk, the 69 $\upmu$m feature rapidly increases in strength and should have been detectable. Because the radial size of the  T $\gtrsim 1000$ K zone, where forsterite can form by annealing or vapor condensation, is small, the presence of forsterite features at short wavelengths imply that forsterite has somewhat diffused outward. However, the absence of the 69 $\upmu$m features indicates that the diffusion process does not efficiently transport forsterite grains to the outer disk farther than $\sim$5 AU. 

The presence of (weak) 69 $\upmu$m features in HD\,104237 and HD\,144668 and their absence in the other flat disks may reflect optical-depth effects. Among all flat disks, HD\,104237 and HD\,144668 have the lowest mm brightness and have the lowest $I_{23}/I_{69}$ (apart from HD\,144432, but this is an upper limit). Since mm emission is optically thin, the low mm luminosities may imply lower disk masses. When the disk mass is lower, the disk is also more optically thin. In that case, the region where the forsterite is located may become optically thin at 69 $\upmu$m while it is still optically thick at MIR wavelengths. Alternatively, forsterite grains may have been diffused outward to colder regions of the disk, or may have grown to larger sizes. All of these effects enhance the strength of the 69 $\upmu$m feature and lower the $I_{23}/I_{69}$ ratio. Although we cannot determine which evolutionary model is favourable, they all suggest that the detection of the 69 $\upmu$m feature in flat disks is connected with disk evolution.

\subsection{Forsterite in transitional disks}

Many if not all Herbig disks with disk classification group I are disks with large gaps and can be characterised as (pre-)
transitional. An evolutionary path from the observed group I to the observed group II sources seems no longer likely. Instead, both might derive from a common ancestor, or group I objects may originate from group II sources \citep{2013Maaskant}. The variety in the results of our study for transitional disks (group I sources) is puzzling, in particular in connection with the fact that cold forsterite appears to be absent in flat disks (group II sources).  For the  transitional disks, we found that the majority of these disks do not have a detectable amount of cold forsterite (measured by the presence or absence of the 69 $\upmu$m feature).  However, for a few sources, significant amounts of forsterite are present in the cold outer regions. The high crystalline mass fraction in the inner edge of the outer disk  of HD\,100546 \citep{2011Mulders} seems to indicate that the cold forsterite is connected to the formation of the gap. As suggested by \citet{2003Bouwman}, forsterite grains may be blown outwards on the wall by radiation pressure. However, this scenario would not work efficiently if significant amounts of gas resided in the gaps (as is the case for the transitional disks of Oph IRS 48, HD\,141569 and HD\,100546, \citealt{2014Bruderer,2014Thi, 2006AckeAncker}). For HD\,100546, the non-negligible accretion rate derived by \citet{2004Deleuil} also argues against outward radial mixing through the gap.  

A possibility to explain the cold forsterite would be that the mixing in an earlier evolutionary state is more efficient than we derived, and that the forsterite is hidden in the disk and only exposed and made visible when the gap opens and the disk rim is illuminated. Perhaps the forsterite grains hide in the midplane after growing and settling and become visible when the disk mass decreases because of optical-depth effects. Another option is that the forsterite is locally produced in the transitional disks where it is observed. This would also suggest a connection to the gap-forming process, but is clearly not a process that is always linked to the formation of gaps. 

When we consider this result in terms of an evolutionary picture, the following scenarios and questions arise: if we assume that transitional disks are the evolutionary successor to flat disks, then we would in the first instance not expect a significant amount of forsterite in the outer disk.  Since radial mixing does not appear to be effective in flat disks, the situation just before the opening of a big gap would be that little or no forsterite is present at low temperatures in the outer disk.  Direct radial mixing would become impossible in a disk with a large gap. If, however, the transitional disks and flat disks have evolved from a common ancestor, but represent different paths of disk evolution, the large dispersion in abundances of cold forsterite in transitional disks still remains to be explained in this framework.

\subsection{Larger grains in group Ib transitional disks}
\label{sec:largergrains}
HD\,141569 and Oph IRS 48 are transitional disk that lack the 10 and 20 $\upmu$m amorphous silicate features, but instead show prominent PAH emission bands in their spectra. This is an indicator of large gaps that have depleted the population of small ($\lesssim$ 1 $\upmu$m) dust grains in the temperature region of the disk where amorphous silicate features are produced \citep{2013Maaskant}. The 69 $\upmu$m features of HD141569 and Oph IRS 48 are remarkably strong and have large widths, indicating temperatures of $\sim150-200$ K. This is inconsistent with the temperature derived from the $I_{23}/I_{69}$ ratio ($\sim50-100$ K), however. We found by detailed modelling of HD\,141569 that there are two scenarios that fit the forsterite spectra.  In the first scenario (model B), the forsterite is located in the outer disks at $\sim200$ AU where temperatures are below $\lesssim40$ K, an iron distribution was used to fit the width of the 69$\upmu$m feature. In the second scenario (model C), the forsterite was modelled by larger ($10$ $\upmu$m) grains in the disk zone that also produces the mid- to far-infrared continuum emission. The larger grain size broadens the 69 $\upmu$m feature and decreases the strength of the features at lower wavelength. Combinations of these two extreme models may also fit the spectra.

We now discuss the forsterite formation histories required to explain the two model solutions. For model B, it is difficult to understand that there is a mass abundances of $\sim$20 \% in the disk at $\sim200$ AU, while no forsterite grains are present in the innermost disk zones. If shock-heating, parent-body collisions, or radial mixing are responsible for the cold forsterite, we may expect this to be more efficient closer to the star because
of higher densities and relative velocities.  The Gaussian distribution of iron fractions between $0-1$~\% has no clear origin either. The shape of the 69 $\upmu$m feature of HD\,141569 is similar to that of Oph IRS 48. The inner edge of the outer disk of Oph IRS 48 has a temperature $\sim$120 K and is located at $\sim$60 AU \citep{2013Maaskant}. However, there is no observational evidence as yet of a large colder (T $\lesssim50$ K) outer disk at $\gtrsim200$ AU for Oph IRS 48. A model with a dominant contribution of cold, iron-rich forsterite (model B) seems much more difficult to construct for Oph IRS 48. Model C, the scenario with a dominant contribution of larger (10 $\upmu$m) grains, does not have these difficulties. In addition, our solution of larger forsterite grains in the disks of HD\,141569 and Oph IRS 48 seems to be consistent with significant depletion of small grains in group Ib sources. For all these reasons, we consider model C to be a more likely scenario to explain the forsterite spectrum of HD\,141569 and Oph IRS 48.  In this scenario, the larger width of the forsterite 69 $\upmu$m feature can be used as an independent indicator of larger grain sizes in protoplanetary disks.


\section{Conclusions}
We have performed a detailed radiative transfer study to examine the observational behaviour of forsterite spectra under several evolutionary scenarios. In particular, we checked for consistency between the forsterite   $I_{23}/I_{69}$ feature strength ratio and the shape of the 69$\upmu$m feature and studied HD\,141569 in detail because it is the most extreme outlier in the sample. Our findings are as follows:

\label{sec:conclusions}
\begin{itemize}
\item Forsterite in flat (group II) disks is located in the inner few AU. This indicates that radial mixing is an inefficient process, at least in flat disks, or that local production does not take place. 
\item The detection rate of the 69 $\upmu$m forsterite feature is higher for objects with lower millimetre luminosity. This may indicate that as disks evolve toward lower masses, optical-depth effects or increased production/transportation of forsterite in/to cold regions of the outer disk enhance the strength of the 69 $\upmu$m feature.
\item The $I_{23}/I_{69}$ ratio is an independent forsterite temperature indicator and is most inconsistent with the observed 69 $\upmu$m shapes of HD\,141569 and Oph IRS 48.
\item The large widths of the 69 $\upmu$m features of HD\,141569 and Oph IRS 48 can be explained by forsterite grains with sizes above a few micron, or by a distribution of iron fractions between \mbox{$\sim0-1$~\%}.
\item The weak 23 $\upmu$m feature compared eith the 69 $\upmu$m band strength in the evolved transitional objects HD\,141569 and IRS 48 can be explained by very cold or larger grains. 
\item The innermost disk of HD\,141569, responsible for the continuum emission between $\sim4-14$ $\upmu$m,  does not contain small ($\lesssim1$ $\upmu$m) amorphous and crystalline silicate grains.
\item Radiative transfer models show two solutions to fit the larger width of the forsterite 69 $\upmu$m band, consistently with the low $I_{23}/I_{69}$ feature strength ratio of HD\,141569. A model with iron-rich ($\sim0-1$ \% Fe fraction), $\sim$40 K, 1 $\upmu$m grains (model B) and a model using iron-free, $\sim$100 K, 10 $\upmu$m grains (model C). 
We argue that the model with larger ($\gtrsim10$ $\upmu$m) forsterite seems to be most likely.

\end{itemize}

\begin{acknowledgements}

The authors thank the anonymous referee for his/her excellent comments and constructive feedback. The authors thank Neal Evans, Gwendolyn Meeus, Thomas Henning, Jeroen Bouwman, and the members of the Herschel DIGIT team for useful discussions that improved the analysis. The authors thank Gijs Mulders and Inga Kamp for useful comments on the manuscript. K.M. is supported by a grant from the Netherlands Research School for Astronomy (NOVA).  M.M. acknowledges funding from the EU FP7-2011 under Grant Agreement No 284405.  Studies of interstellar chemistry at Leiden Observatory are supported through advanced-ERC grant 246976 from the European Research Council, through a grant by the Dutch Science Agency, NWO, as part of the Dutch Astrochemistry Network, and through the Spinoza premie from the Dutch Science Agency, NWO.

 \end{acknowledgements}



\end{document}